\documentclass[twocolumn, showpacs, preprintnumbers, amsmath, amssymb,noshowpacs, nopreprintnumbers]{revtex4}
\usepackage{pifont}
\usepackage{tikz}
\usetikzlibrary{shapes,arrows}
\usepackage[colorlinks,urlcolor=blue]{hyperref}
\usepackage{epstopdf}

\newcommand{\reform}{\color{black} }
\usepackage{graphicx}% Include figure files
\usepackage{dcolumn}% Align table columns on decimal point
\usepackage{appendix}
\usepackage{breakurl}%
\usepackage{changes}

\makeatletter
  \let\@font@info\@gobble
  \let\@font@warning\@gobble
\makeatother

\begin{document}

%\preprint{APS/123-QED}

\title{Diffusion Equation and Rare Fluctuations of the Biased ACTRW Model}

\author{Yuanze Hong$^1$}

\author{Tian Zhou$^2$}

 %\altaffiliation[Also at ]{School of Mathematics and Statistics, Lanzhou University.}%Lines break automatically or can be forced with \\

\author{Wanli Wang$^1$}%
\email[contact author: ]{wanliiwang@163.com}
\affiliation{%
%  $^1$School of Mathematics and Statistics, Gansu Key Laboratory of Applied Mathematics and Complex
%Systems, Lanzhou University, Lanzhou 730000,  P.R. China
%\\
$^1$ School of Mathematical Sciences, Zhejiang University of Technology, Hangzhou 310023, China
 }%
\affiliation{$^2$ School of Information Engineering University,
Zhengzhou, Henan 450001, People's Republic of China}

% \homepage{http://www. Second. institution. edu/~Charlie. Author}
%\affiliation{

%\date{\today}% It is always \today,  today,
\begin{abstract}
We explore the fractional advection-diffusion equation and rare events associated with the ACTRW model. When waiting times have a finite mean but infinite variance, and the displacements follow a narrow distribution, the fractional operator is defined in terms of space rather than time. The far tail of the positional distribution is governed by rare events, which exhibit a different scaling compared to typical fluctuations. Additionally, we establish a strong relationship between the number of renewals and the positional distribution in the context of large deviations. Throughout the manuscript, the theoretical results are validated through simulations.
\end{abstract}

%\pacs{02. 50. -r,  05. 20. -y,  05. 40. -a }% PACS,  the Physics and Astronomy
% url
%https://publishing.aip.org/publishing/pacs/pacs-reg00#05
%
%   02. 50. -r   Probability theory, stochastic processes, and statistics
%  05.30.Pr	Fractional statistics systems (anyons, etc.)

% 05.40.-a Fluctuation phenomena, random processes, noise, and Brownian motion (for fluctuations in superconductivity, see 74.40.-n; for statistical theory and fluctuations in nuclear reactions, see 24.60.-k; for fluctuations in plasma, see 52.25.Gj; for nonlinear dynamics and chaos, see 05.45.-a)
%
%??05.10.Gg Stochastic analysis methods (Fokker-Planck, Langevin, etc.)

% 02.50.Ey	Stochastic processes
%  45.10.Hj	Perturbation and fractional calculus methods

\maketitle

\section{Introduction}\label{18ctrwsect1}
The continuous-time random walk (CTRW) model is widely used to describe a variety of phenomena across disciplines such as physics, chemistry, biophysics, and biochemistry. For a comprehensive review, see \cite{Metzler2000random,Klafter2011First}. Over the past two decades, the aging phenomenon has found significant applications in fields ranging from biology to chemistry and materials science. Examples include statistical aging and ergodicity breaking in semiconductor CdSe nanocrystals \cite{Brokmann2003Statistical}, time-ensemble-averaged mean squared displacement (MSD) in studies of diffusion within the plasma membrane of living cells \cite{Weigel2011Ergodic}, anomalous transport in cellular flows \cite{Patrick2016Anomalous}, and neuronal messenger ribonucleoprotein transport \cite{Song2018Neuronal}. One widely used approach for addressing aging is the aging continuous-time random walk (ACTRW) model. Theoretically, aging behavior has been investigated through a deterministic dynamical system within the framework of the aging diffusion equation \cite{Barkai2003Agingin}. {\reform The relation between generalized master equation and ACTRW was given in \cite{Metzler2000Generalized,Allegrini2003Generalized}.}
Time averages of single-particle trajectories demonstrate population splitting and aging effects in scale-free anomalous diffusion processes \cite{Schulz2013Aging}. The self-similarity property of the uncoupled ACTRW was explored in \cite{Busani2014AgingUC}, and for more details on aging renewal theory and its applications, see \cite{Schulz2014Aging}. Furthermore, equilibrium initial conditions have been discussed in the contexts of both renewal processes \cite{Wang2019Transport} and CTRW models \cite{Hou2018Biased}. See also discussions related to aging L{\'e}vy walk \cite{Marcin2017Aging}, {\reform and correlation functions together with aging effects in the context of a two-state statistical process \cite{Allegrini2004Non,Allegrini2005Correlation,Bruce2008Maximizing}}.

Most previous studies have focused on cases where the waiting times have an infinite mean. In such scenarios, the system is strongly influenced by the aging time, as the mean of the first waiting time is infinite. However, the situation changes when waiting times have a finite mean but an infinite variance. In this case, the average of the first waiting time asymptotically scales as $t_a^{2-\alpha}$ \cite{Wang2018Renewal}, where $t_a$ is the aging time of the system. When $t_a \to \infty$, corresponding to equilibrium initial conditions, the average of the first waiting time tends to infinity. On the other hand, for a finite $t_a$, the mean of the first waiting time remains finite. In that sense, observing aging effects in the ACTRW model would be difficult. This manuscript explores these properties in detail.

Fractional kinetic equations are widely used to model anomalous diffusion, ranging from sub-diffusion \cite{Metzler1999Anomalous,Barkai2001Fractional,Magdziarz2008Equivalence,Deng2008Finite,Chechkin2005Fractional} to super-diffusion \cite{Wang2020Fractional,Wang2024Fractional}. {\reform When waiting time has an infinite mean, the memory kernel of the general diffusion equation is with respect to time \cite{Allegrini2003Generalized,Bruce2008Maximizing,Metzler1999Deriving}, which results in a  fractional time operator.  If the waiting times have a finite mean, but an infinite variance, and the displacement follows a narrow distribution with a finite, non-zero mean, the fractional operator applies to the spatial domain rather than the temporal domain for a CTRW model \cite{Wang2020Fractional}, which makes a significant departure from the scenario discussed in \cite{Metzler1998Anomalous,Allegrini2004Non}. The introduction of aging into the system has been shown to result in a sub-diffusion-type equation \cite{Barkai2003Agingin} when waiting times have an infinite mean.
However, when waiting time has a finite mean and an infinite variance, the behavior of the diffusion equation with aging remains unknown, which we address in this manuscript. To be more exactly, we use the subordination method to obtained the positional distribution in the long time limit and the corresponding diffusion equation.
}

The second aim of this manuscript is to investigate rare events of the positional distribution. These rare events are linked to large deviation theory, which focuses on the tails of distributions of observables, as opposed to typical fluctuations that characterize the central region of the PDF. For instance, recently, one of the hot topics in large deviations theory is the exponential decay investigated in \cite{Pinaki2007Universal, Barkai2020Packets, Wang2020Large, Burov2022Exponential}.  This work concentrates on the statistics of positional distribution within the long-time limit. Recall that the large deviations in the far tail for $t_a \to \infty$ were previously explored in \cite{Wang2019Transport}; here, we extend this to a more general case where $t_a$ is finite.

We further investigate the relationship between the positional distribution and the number of renewals, exploring the strong connection between these two quantities. We focus on small values of the position $x$, which stems from extremely long trapping times. Similarly, long sojourn times lead to a small number of renewals in the context of renewal processes. Therefore, we expect a correlation between small positional values and a low number of renewals. This spirit aligns with the single big jump principle discussed in \cite{Alessandro2019Single,Wang2019Transport}, as applied to L{\'e}vy walks and the CTRW model. In this sense, rare events of one observable can provide insights into another.

The paper is organized as follows. In Sec. \ref{CTRWANDActrw}, we introduce the CTRW and ACTRW models, including the specific case of the equilibrium process. In Sec. \ref{AFADAE}, we derive the aging fractional advection diffusion  equation using the subordination approach. Sec. \ref{Strongrelation} explores rare fluctuations in the positional distribution and the number of renewals. Finally, a summary is provided in Sec.\ref{Conclusion}.

\section{CTRW and ACTRW}\label{CTRWANDActrw}

The CTRW model, originally proposed by Montroll and Weiss, is a widely applicable framework in the study of advection-diffusion processes. The model is defined as follows: A particle starts at the initial position $x = 0$ at time $t = 0$. The particle remains at this position for a random duration $\tau_1$, drawn from the probability density function (PDF) $\phi(\tau)$, before making a jump to a new position $x_1$, determined by the PDF $f(\chi)$. After arriving at $x_1$, the particle stays there for $\tau_2$, then jumps again to a new position $x_1 + x_2$ at time $t = \tau_1 + \tau_2$. The process is then renewed, and terminated at some time $t$. The CTRW model assumes a specific set of initial conditions known as non-equilibrium initial conditions. In this model, all the waiting times $\tau_i$ and displacements $x_i$ are mutually independent and identically (IID) distributed random variables, governed by $\phi(\tau)$ and $f(\chi)$, respectively.

However, in the real world, determining the exact starting point of a process can be challenging. For instance, it may be difficult to identify when observable, such as molecules in a cell, begin to move. To address this, the ACTRW model is introduced \cite{Monthus1996Models,Barkai2003Aging,Schulz2014Aging,Wang2017Aging}. The key difference between CTRW and ACTRW lies in the modification of the first waiting time $\tau_1$. Specifically, for ACTRW, the process is considered to start at time $t = -t_a$ rather than $t = 0$, and the process is observed at time $t=0$. As a result, generally, the waiting time for the first step depends on $t_a$, reflecting aging effects. Let $\omega(t_a, t)$ be the PDF of the first waiting time. In the double Laplace space, with $ t\to s$ and $t_a\to u$, this can be expressed as follows \cite{Godreche2001Statistics,Schulz2014Aging,Wang2018Renewal}
\begin{equation}\label{firtwaitingtime}
\widehat{\omega}(u,s)= \frac{\widehat{\phi}(u)-\widehat{\phi}(s)}{s-u}\frac{1}{1-\widehat{\phi}(u)}.
\end{equation}
Here, $\widehat{\phi}(s)=\mathcal{L}[\phi(\tau)]=\int_0^\infty \exp(-s\tau)\phi(\tau)d\tau$ is the Laplace transform of $\phi(\tau)$. Notably, when $t_a \to \infty$, the process is referred to as an equilibrium CTRW, provided that $\langle\tau\rangle = \int_0^\infty \tau \phi(\tau) \, d\tau$ is finite. It was discussed previously in the context
of renewal process and random walk \cite{Aquino2010Beyond,Aquino2011Transmission,Hou2018Biased,Wang2019Transport}.  In this limit, $\omega(t_a, t)$ simplifies to
\begin{equation}\label{eqPower}
\begin{split}
 \lim_{t_a\to\infty}\omega(t_a,t) & =\frac{\int_t^\infty \phi(y) \, dy}{\langle \tau\rangle} \\
  & = g(t).
\end{split}
\end{equation}
Here, we assign $g(t)$ as the limiting distribution of $\omega(t_a, t)$ for $t_a\to\infty$.

Throughout this manuscript, we consider a widely applicable case characterized by a fat-tailed distribution \cite{Metzler2000random,Levy2003Measurement,Aquino2010Beyond,Aquino2011Transmission,Burioni2013Rare,Alon2017Time,Deng2020Modeling,Tian2020Continuous,Bodrova2020Continuous,Liu2023Levy}
\begin{equation}\label{powerlaw}
\phi(\tau)= \begin{cases}0, & \tau\leq\tau_0 ;\\ \alpha \frac{\tau_{0}^{\alpha}}{\tau^{1+\alpha}}, & \tau>\tau_{0} .\end{cases}
\end{equation}
Here, $\alpha > 0$ is the index of the power-law distribution, and $\tau_0$ is the cutoff. When $0 < \alpha < 1$, the waiting times have an infinite mean. In this work, we concentrate on the widely applicable case where $1 < \alpha < 2$ \cite{Margolin2002Spatial,Levy2003Measurement,Schroer2013Anomalous,Liang2015Sample,Alon2017Time,Wang2020Strong,Zhang2022Correlated}. In this range, the waiting times have a finite mean but an infinite variance. The Laplace transform of Eq.~\eqref{powerlaw} can be expressed by
\begin{equation}\label{LapPower}
\widehat{\phi}(s) \sim 1 - \langle \tau \rangle s + b_{\alpha} s^{\alpha},
\end{equation}
where $b_{\alpha} = \tau_0^{\alpha} |\Gamma(1-\alpha)|$ and $s \to 0$. The leading term in Eq.~\eqref{LapPower}, represented by $1$, ensures the normalization condition, i.e., $\widehat{\phi}(0) = 1$. The symbol $\langle \tau \rangle$ represents the mean waiting time. The term $s^\alpha$ stems from the heavy tail of Eq.~\eqref{powerlaw}.

Assume that the PDF of jump step length $\chi$ obeys
\begin{equation}\label{Guassian}
f(\chi)=\frac{1}{\sqrt{2\sigma^{2}\pi}}\exp\left[-\frac{(\chi -a)^{2}}{2\sigma^{2}}\right]
\end{equation}
with mean $a>0$ and the variance $\sigma^2$. Applying the Fourier transform on Eq.~\eqref{Guassian}, we obtain $\widetilde{f}(k)=\exp (ika-\sigma^{2}k^{2}/2)$ with $\widetilde{f}(k)=\mathcal{F}[f(\chi)]=\int_{-\infty}^\infty \exp(-ik\chi)f(\chi)d\chi$. In the limit of $k\to 0$, the Taylor expansion yields
\begin{equation}\label{GaussianF}
  \begin{aligned}
    \widetilde{f}(k)\sim 1+ika-\frac{\sigma^{2}+a^{2}}{2}k^{2}.
  \end{aligned}
\end{equation}
It is important to note that our results discussed below are not limited to Eq.~\eqref{Guassian}; the general condition is that the displacements have a finite, non-zero mean and a finite variance. From Eq.~\eqref{eqPower}, when $1 < \alpha < 2$, the first waiting time $\tau_1$ has an infinite mean. If $t_a$ is not very large, the impact of aging on the statistics of the random walk and renewal process is interesting and warrants careful consideration.

\section{Aging Fractional Advection Diffusion Asymmetry Equation}\label{AFADAE}
We now proceed to derive the aging fractional advection-diffusion asymmetry equation. By incorporating appropriate initial or boundary conditions, the kinetic equations provide a straightforward approach that simplifies the more complex CTRW model, enabling the solution of more intricate dynamics.

Let $P(x, t_a, t)$ represent the green function of the ACTRW
\begin{equation}\label{greenF}
P(x, t_a,t) = \sum_{N=0}^\infty Q_{t_a,t}(N) f(x|N),
\end{equation}
where $Q_{t_a,t}(N)$ denotes the probability of observing $N$ renewals between time zero and $t$ for an aging time $t_a$, and $f(x|N)$ is the conditional probability of being at position $x$ given that exactly $N$ steps have occurred at the observation time $t$. Eq.~\eqref{greenF}, known as the subordination formula, is widely applied in physics and mathematics, ranging from fractional kinetic equations \cite{Bouchaud1990Anomalous,Klafter1994Probability,Barkai2001Fractional,Wang2020Fractional,Trifce2024Fractional} to financial models \cite{semeraro2022multivariate}.

In double Laplace spaces, where $t \to s$ and $t_a \to u$, $Q_{t_a,t}(N)$ can be expressed as follows
\begin{equation}\label{QUS}
\widehat{Q}_{u,s}(N)= \begin{cases}
\frac{1}{su}-\frac{\widehat{\omega}(u,s)}{s}, & N=0;\\  \widehat{\omega}(u,s)\widehat{\phi}^{N-1}(s)\dfrac{1-\widehat{\phi}(s)}{s}, & N\geq 1 \end{cases}
\end{equation}
using renewal theories. The details of Eq.~\eqref{QUS} can be found in \cite{Barkai2003Agingin,Schulz2014Aging}. In this work, we focus on the statistics of large $N$ in the long-time limit, which leads to the relation
\begin{equation}\label{QusMving}
\widehat{Q}_{u,s}(N)\sim\widehat{\omega}(u,s)\widehat{Q}_s(N),
\end{equation}
where \begin{equation}\widehat{Q}_s(N) = \widehat{\phi}^{N}(s)\dfrac{1-\widehat{\phi}(s)}{s}
\end{equation} describes the PDF of the number of renewals without aging \cite{Godreche2001Statistics}, i.e., the probability of observing $N$ renewals at the observation time $t$ for the normal renewal process. Mathematically, from Eq.~\eqref{QusMving}, $Q_{t_a,t}(N)$ can be expressed as
\begin{equation}\label{QtatC}
Q_{t_a,t}(N)\sim\int_0^t\omega(t_a,t-y)Q_{y}(N)dy
\end{equation}
using the double inverse Laplace transforms.
For typical fluctuations of $N$, the scaling when $N - t/\langle \tau \rangle \sim t^{1/\alpha}$ gives
\begin{equation}\label{QtN}
Q_t(N)\sim \frac{1}{(t/\bar{t})^{1/\alpha}}\mathcal{L}_{\alpha}\left(\frac{N-t/\langle\tau\rangle}{(t/\bar{t})^{1/\alpha}}\right),
\end{equation}
where $\bar{t} = \langle \tau \rangle^{1+\alpha}/(\tau_0^\alpha |\Gamma(1-\alpha)|)$. The function $\mathcal{L}_{\alpha}(x)$ represents the asymmetric L{\'e}vy stable distribution, with $\mathcal{F}[\mathcal{L}_{\alpha}(x)] = \exp((-ik)^\alpha)$. Note that Eq.~\eqref{QtN} demonstrates an asymmetric behavior of $N$ characterized by a fat tail and a narrow one.

\begin{figure}[htp]
\centerline{\includegraphics[width=22pc]{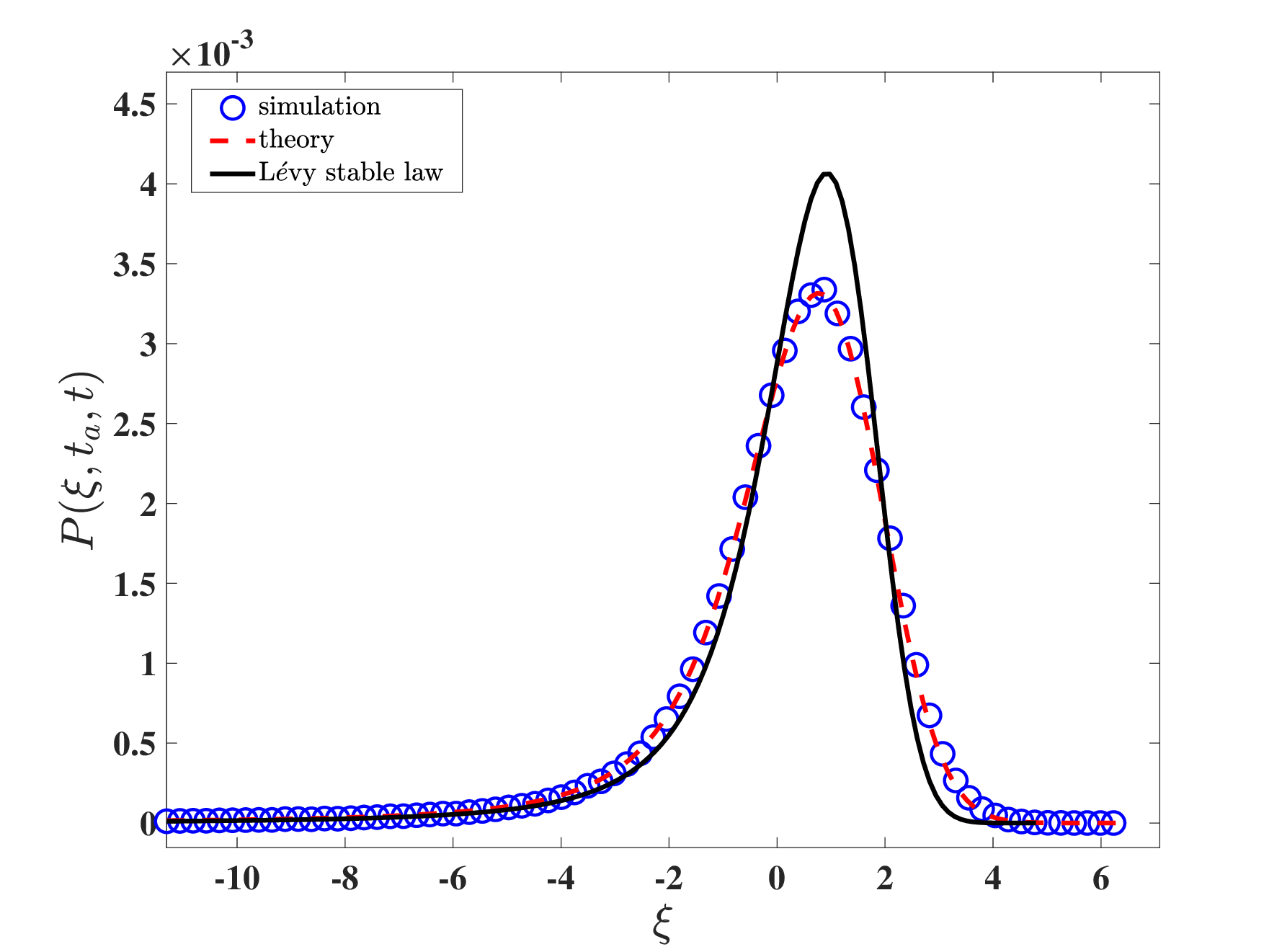}}
  \caption{Comparison  Eq.~\eqref{ffeq} with the L{\'e}vy stable law Eq.~\eqref{LevylAW} with the scaling $\xi =(x-at/\langle\tau\rangle)/(a(t/\bar{t})^{1/\alpha})$. The theory plotted by the dashed line is the solution of the aging fractional advection-diffusion equation given in Eq.~\eqref{soluEEq} and the L{\'e}vy stable law shows typical fluctuations of the position obtained from Eq.~\eqref{LevylAW} describing typical fluctuations of the position. We use the parameters $\alpha=3/2$, $\tau_0 = 0.1$, $t = 1000$, $t_a=1000$, $a=1/2$, and $\sigma = 1$. }\label{LevyLaw}
\end{figure}

When the aging time $t_a $ tends to infinity  and the observation time $ t $ is long, the positional distribution converges to the well-known L{\'e}vy stable law, as derived using Eqs.~\eqref{LapPower}, \eqref{GaussianF}, \eqref{greenF}, and \eqref{QUS}
\begin{equation}\label{LevylAW}
P(x,t_a,t) \sim \frac{1}{a (t/\bar{t})^{1/\alpha}} \mathcal{L}_{\alpha}\left(\frac{\frac{x}{a} - \frac{t}{\langle \tau \rangle}}{(t/\bar{t})^{1/\alpha}}\right).
\end{equation}
This result can also be obtained from Eq.~\eqref{QtN} using the relation $x \sim aN$. It can be seen that Eq.~\eqref{LevylAW} is independent of the aging time $ t_a$ of the system.
In our simulations, the convergence to Eq.~\eqref{LevylAW} is observed to be suboptimal, as shown by the solid line in Fig.\ref{LevyLaw}. To address this, we propose an enhancement using a fractional advection-diffusion equation, which provides a more accurate description of the positional distribution.

In the following, we treat $t_a$ as a parameter without specifying whether it is large or small, while assuming that $t$ is sufficiently long.  Using Eqs.~\eqref{greenF} and \eqref{QtatC}, we obtain
\begin{equation}\label{PxtLong}
\begin{split}
    P(x,t_a,t)&\sim\int_0^{\infty}  \frac{\exp \left(-\frac{(x-a N)^2}{2 \sigma^2 N}\right)}{\sqrt{2 \pi \sigma^2 N}} \\&\times\int_0^t\frac{\omega(t_a,t-y)}{(y/\bar{t})^{1/\alpha}}\mathcal{L}_{\alpha}\left(\frac{N-y/\langle\tau\rangle}{(y/\bar{t})^{1/\alpha}}\right)dydN,
\end{split}
\end{equation}
where we have used the fact that displacements are IID random variables drawn from Eq.~\eqref{Guassian}. Changing the variables $\xi = (N-y/\langle \tau \rangle)/(y/\bar{t})^{1/\alpha}$, Eq.~\eqref{PxtLong} can be further rearranged
\begin{equation}
\begin{split}
    P(x,t_a,t)&\sim\int_0^{t}\omega(t_a,t-y)  \int_{-\frac{y}{\langle\tau\rangle}(\frac{y}{\bar{t}})^{\frac{1}{\alpha}}}^\infty \mathcal{L}_{\alpha}(\xi)\\
    &\times\frac{\exp \left(-\frac{(x-a (\xi(\frac{y}{\bar{t}})^{\frac{1}{\alpha}}+\frac{y}{\langle\tau\rangle}))^2}{2 \sigma^2 (\xi(\frac{y}{\bar{t}})^{\frac{1}{\alpha}}+\frac{y}{\langle\tau\rangle})}\right)}{\sqrt{2 \pi \sigma^2 (\xi(\frac{y}{\bar{t}})^{\frac{1}{\alpha}}+\frac{y}{\langle\tau\rangle})}}d\xi dy.
\end{split}
\end{equation}
The above equation can be used to predict the positional distribution of the ACTRW model.
In the long-time limit, by replacing $\frac{y}{\langle \tau \rangle} \left( \frac{y}{\bar{t}} \right)^{\frac{1}{\alpha}}$ with infinity, the above expression simplifies to
\begin{equation}
\begin{split}
    P(x,t_a,t)&\sim\int_0^{t}\omega(t_a,t-y)  \int_{-\infty}^\infty \mathcal{L}_{\alpha}(\xi)\\
    &\times\frac{\exp \left(-\frac{(x-a (\xi(\frac{y}{\bar{t}})^{\frac{1}{\alpha}}+\frac{y}{\langle\tau\rangle}))^2}{2 \sigma^2 (\xi(\frac{y}{\bar{t}})^{\frac{1}{\alpha}}+\frac{y}{\langle\tau\rangle})}\right)}{\sqrt{2 \pi \sigma^2 (\xi(\frac{y}{\bar{t}})^{\frac{1}{\alpha}}+\frac{y}{\langle\tau\rangle})}}d\xi dy.
\end{split}
\end{equation}
Taking the Fourier transform concerning $x$, we obtain
\begin{equation}
\begin{split}
    \widetilde{P}(k,t_a,t)&\sim\int_0^{t}\omega(t_a,t-y)  \int_{-\infty}^\infty \mathcal{L}_{\alpha}(\xi)\\
    &~~~\times \exp((-ika-\frac{1}{2}\sigma^2k^2)\\
    &~~~\times [(\xi(\frac{y}{\bar{t}})^{\frac{1}{\alpha}}+\frac{y}{\langle\tau\rangle})])d\xi dy.
\end{split}
\end{equation}
Rearranging the above expression, we have
\begin{equation}
\begin{split}
    \widetilde{P}(k,t_a,t)&\sim\int_0^{t}\omega(t_a,t-y) \\
    &\times\exp((-ika-\frac{1}{2}\sigma^2k^2)\frac{y}{\langle\tau\rangle}) \int_{-\infty}^\infty \mathcal{L}_{\alpha}(\xi)\\
    &\times \exp((-ika-\frac{1}{2}\sigma^2k^2)(\xi(\frac{y}{\bar{t}})^{\frac{1}{\alpha}}))d\xi dy.
\end{split}
\end{equation}
In the long-time limit, where $x \to \infty$, the term $\frac{1}{2}\sigma^2 k^2$ in the third line can be neglected as $k \to 0$. Thus, we obtain
%\begin{equation}
%\begin{split}
 %   \widetilde{\mathcal{P}}(k,t_a,t)&=\int_0^{t}\omega(t_a,t-y) \\
  %  &\times\exp((-kia-\frac{1}{2}\delta^2k^2)\frac{y}{\langle\tau\rangle}) \int_{-\infty}^\infty \mathcal{L}_{\alpha}(\xi)\\
  %  &\times \exp((-kia)[(\xi(\frac{y}{\bar{t}})^{\frac{1}{\alpha}})])d\xi\mathrm{d}y
%\end{split}
%\end{equation}
%or
\begin{equation}\label{fffxo1}
\begin{split}
    \widetilde{\mathcal{P}}(k,t_a,t)=&\int_0^{t}\omega(t_a,t-y)\\
    &\times\exp((-ika-\frac{1}{2}\sigma^2k^2)\frac{y}{\langle\tau\rangle})\\
    &\times \exp((-ika)^\alpha \frac{y}{\bar{t}})dy,
\end{split}
\end{equation}
where we denote $\widetilde{\mathcal{P}}(k,t_a,t)$ or $\mathcal{P}(x,t_a,t)$ as the limiting law of $\widetilde{P}(k,t_a,t)$ or $P(x,t_a,t)$ in the long time limit.
Rewriting the above equation, we have
\begin{equation}\label{pktatwiw}
\begin{split}
    \widetilde{\mathcal{P}}(k,t_a,t)&=\int_0^{t}\omega(t_a,y)\\
    &\times \exp((-ika-\frac{1}{2}\sigma^2k^2)\frac{t-y}{\langle\tau\rangle})\\
    &\times \exp((-ika)^\alpha \frac{t-y}{\bar{t}})dy.
\end{split}
\end{equation}
%Let $k=0$, $\widetilde{\mathcal{P}}(k=0,t_a,t)=\int_0^t\omega(t_a,y)dy\sim 1$. It demonstrates that $\mathcal{P}(x,t_a,t)$ is normalized.
Taking the derivative of this equation with respect to time yields
\begin{equation*}
\begin{split}
\frac{\partial \widetilde{\mathcal{P}}(k,t_a,t)}{\partial t}&=[(-ika-\frac{1}{2}\sigma^2k^2)\frac{1}{\langle\tau\rangle}+\frac{a^\alpha}{\bar{t}}(-ik)^\alpha]\\
&~~~\times \widetilde{\mathcal{P}}(k,t_a,t)+\omega(t_a,t).
\end{split}
\end{equation*}
The inverse Fourier transform of the above equation gives the main result of this section
\begin{equation}\label{ffeq}
  \begin{aligned}
    \frac{\partial \mathcal{P}}{\partial t}= & D\frac{\partial^2\mathcal{P}}{\partial x^2}-V \frac{\partial \mathcal{P}}{\partial x}+ S\frac{\partial^\alpha \mathcal{P}}{\partial (-x)^\alpha}+\omega(t_a,t)\delta(x)
  \end{aligned}
\end{equation}
with transport constants
\begin{equation}\label{ffeqcon}
D=\frac{\sigma^2}{2\langle\tau\rangle},~~V=\frac{a}{\langle\tau\rangle},~~ S=\frac{a^\alpha}{\bar{t}}.
\end{equation}
The first two constants, $D$ and $V$, are standard terms in the classical advection-diffusion equation. The last two constants are bias-dependent; in particular, the last constant $S$ controls the asymmetric fat-tail of the positional distribution. The operator $\partial^\alpha/\partial (-x)^\alpha$ is known as the right Riemann-Liouville derivative operator with $1<\alpha<2$ \cite{Oldham1974Fractional,Benson2000fractional-order}. In Fourier space, the fractional operator has a simple expression: $\mathcal{F}\left[\partial^\alpha/\partial (-x)^\alpha g(x)\right] = (-ik)^\alpha \widetilde{g}(k)$. The last term, the source term, on the right-hand side of Eq.~\eqref{ffeq}, is the aging-dependent term describing the non-moving particles. If the singular term is ignored, Eq.~\eqref{ffeq} reduces to the fractional advection-diffusion asymmetric equation for the CTRW model \cite{Wang2020Fractional}. Furthermore, Eq.~\eqref{ffeq} indicates that a fat tail of the displacement distribution is not necessary for the emergence of a fractional space operator, as discussed in \cite{Wang2020Fractional}.

The method presented in this manuscript can also be applied to the case of $0 < \alpha < 1$, as explored in \cite{Barkai2003Agingin}, where waiting times possess an infinite mean. In this context, the fractional operator $(-ik)^\alpha$ is replaced by the fractional time Riemann-Liouville operator \cite{Oldham1974Fractional,Podlubny1999Fractional,Metzler2000random,Barkai2001Fractional,Deng2018High}.

\begin{figure}[htp]
\centerline{\includegraphics[width=22pc]{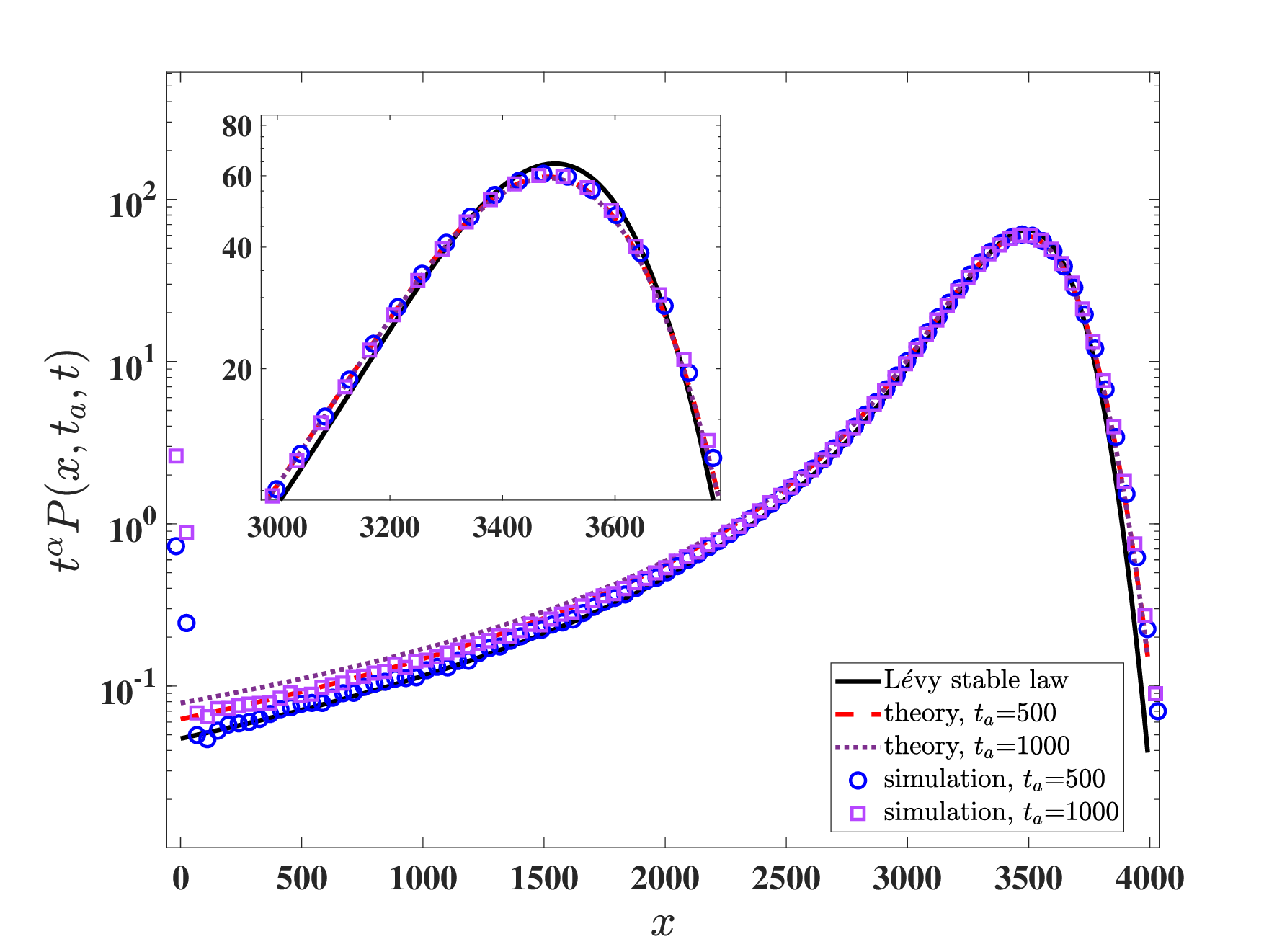}}
\caption{Statistics of the positional for different aging times with a strong bias. Theoretical predictions, plotted by the dashed and dotted lines, are solutions of the aging fractional advection-diffusion equation given in Eq.~\eqref{soluEEq} and the solid line obtained from Eq.~\eqref{LevylAW}. We use the parameters $\alpha=3/2$, $\tau_0 = 0.1$, $t = 1000$, $t_a=1000$, $a = 1$, and $\sigma = 1$. }\label{not_sensitive}
\end{figure}

By the convolution properties of the Fourier transform, the solution of the fractional equation \eqref{ffeq} follows
\begin{equation}\label{soluEEq}
\begin{aligned}
\mathcal{P}(x,t_a,t)=&\omega(t_a,t)\otimes_t\left[\left(\frac{1}{(t/\bar{t})^{\frac{1}{\alpha}}}\right)^{\frac{1}{\alpha}}\mathcal{L}_{\alpha}\left(\frac{x}{(t/\bar{t})^{\frac{1}{\alpha}}}\right)\right.\\
    &\otimes_x\left.\frac{\exp \left(-\frac{(\langle\tau\rangle x-at)^2}{2 t\langle\tau\rangle\sigma^2 }\right)}{\sqrt{2 \pi \sigma^2 t/\langle\tau\rangle}}\right],
\end{aligned}
\end{equation}
where $\otimes_t$ and $\otimes_x$ are the convolution of $t$ and $x$, respectively. In Fig.~\ref{LevyLaw}, Eq.~\eqref{soluEEq} is compared with the asymmetric L{\'e}vy stable law, showing an excellent agreement with the simulations. In our comparison, we find that Eq.~\eqref{soluEEq} is not sensitive to the aging time $t_a$ as shown in Fig.~\ref{not_sensitive} when the time is long and the bias is strong.

\begin{figure}[htp]
\centerline{\includegraphics[width=22pc]{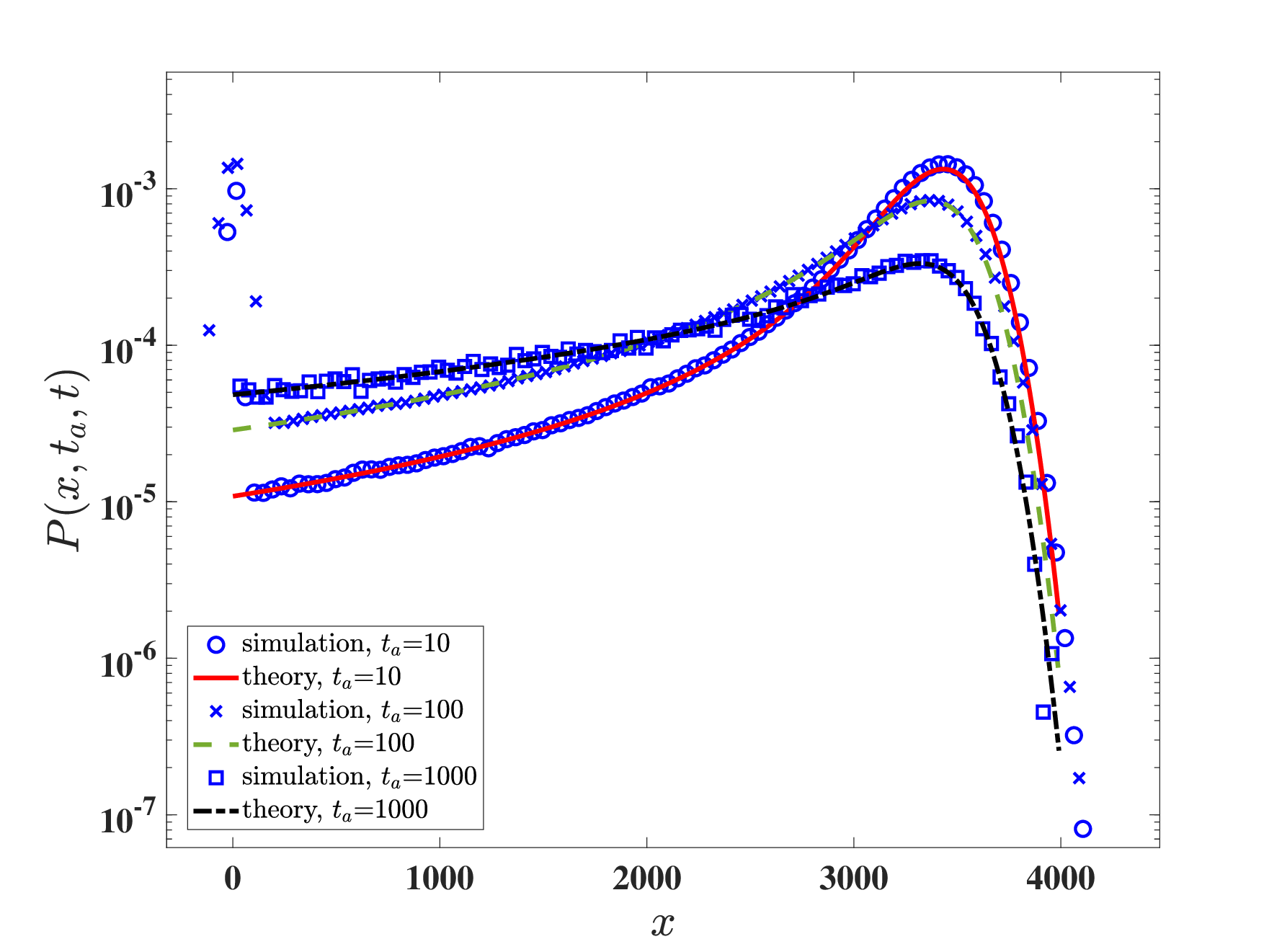}}
  \caption{Application of Eq.~\eqref{ffeq} to a modified ACTRW model, with $t_a = 10$, $100$, and $1000$. We use the parameters $\alpha_1 = 1/2$, $\alpha_2 = 3/2$, $\tau_0 = 0.1$, $t = 1000$, $a = 1$, and $\sigma = 1$. The lines represent the analytical solution from Eq.~\eqref{soluEEq}, while the simulation results are obtained from particle realizations.
  For Eq.~\eqref{soluEEq}, the PDF of the first waiting time follows $\omega(t_a,t)=\sin(\pi\alpha_1)/(\pi(t_a+t)(t/t_a)^{\alpha_1})$ for $t\geq \tau_0$ and $\omega(t_a,t)=\int_t^{\infty}\phi(\tau)d\tau/(b_{\alpha_1}\Gamma(\alpha_{1})t^{1-\alpha_1})$ for $t\leq \tau_0$; see the derivation discussed in \cite{Wang2018Renewal}.
  }
  \label{ComplexAgingFokker}
\end{figure}

To further validate Eq.~\eqref{ffeq}, we consider a more mathematically complex model, i.e., a modified ACTRW model. In this model, during the time interval $(-t_a,0)$, waiting times are generated from Eq.~\eqref{powerlaw} but with an infinite mean, i.e., $0<\alpha =\alpha_1<1$, while in the time interval $(0,t)$,  waiting times are drawn from Eq.~\eqref{powerlaw} with $1<\alpha=\alpha_2<2$. All along the process, the displacements are drawn from Eq.~\eqref{Guassian}.
Figure \ref{ComplexAgingFokker} indicates that Eq.~\eqref{ffeq} provides an accurate prediction.

Using the relationship between the characteristic function and moments, Eq.~\eqref{fffxo1} yields
\begin{equation}
\begin{split}
\langle x(t)\rangle &\sim \frac{a}{\langle\tau\rangle} \int_0^t\omega(t_a, y) (t-y)dy \sim a \frac{t}{\langle\tau\rangle}
\end{split}
\end{equation}
since the mean of the forward recurrence time follows $t^{2-\alpha}$ with  $1<\alpha<2$.
The above equation shows that the aging fractional advection-diffusion equation provides a reasonable prediction for the mean of the ACTRW model. However, for the MSD, Eq.~\eqref{ffeq} fails, as the second moments of the positions obtained from Eq.~\eqref{ffeq} are infinite. Clearly, this prediction is wrong. Besides, when the observation time is long, and the bias is strong, there is no big gap for different aging times in the context of the central part of the positional distribution. An interesting question is whether there is a way to distinguish between them.
These issues will be addressed below using the large deviations theory.

\section{Rare events of the position}
In this section, we focus on large deviations of the position distribution, specifically analyzing the scaling behavior when $x - at/\langle\tau\rangle$ is on the order of $t$. This scaling deviates from the solution of Eq.~\eqref{ffeq}, which primarily describes the central part of the position distribution. We will demonstrate that the far tail of the distribution governs the position's MSD.

\begin{figure}[htp]
  \centerline{\includegraphics[width=22pc]{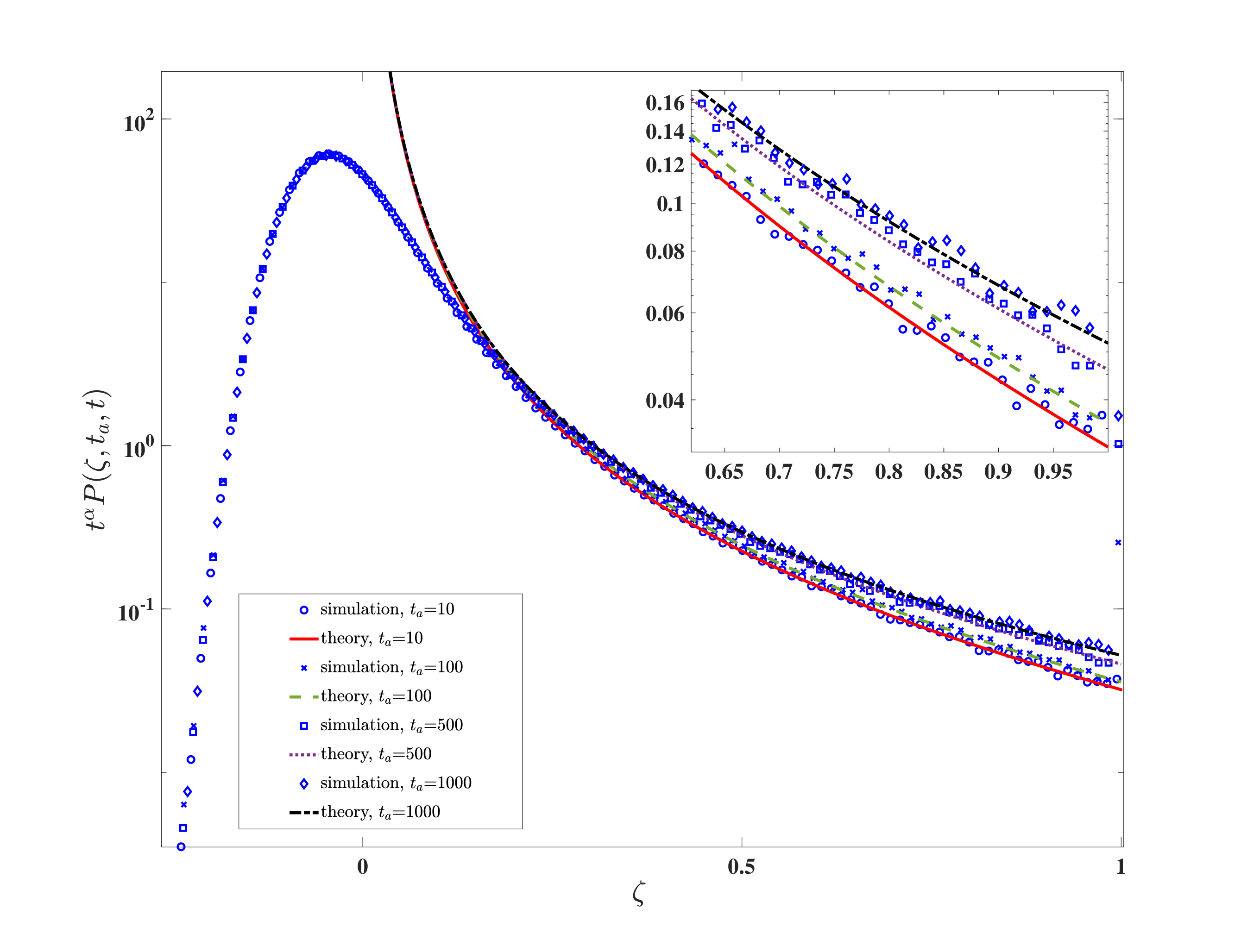}}
  \caption{The scaled PDF describing the far tail of position with $\zeta = 1 - (x/a)/(t/\langle\tau\rangle)$. Symbols represent simulation results based on $10^7$ realizations, while the lines depict the theoretical predictions from Eq.~\eqref{endPxtta1}. The parameters are $\alpha = 3/2$, $a = 1$, $\sigma = 1$, $t = 1000$, $\tau_0 = 0.1$, and $t_a = 10, 100, 500$, and $1000$.}
  \label{RareEVENTSofX}
\end{figure}

Using Eqs.~\eqref{greenF} and \eqref{QUS}, the positional distribution $\widetilde{\widehat{P}}(k,u,s)$ in Fourier-Laplace spaces is given by \cite{Barkai2003Aging,Barkai2003Agingin}
\begin{equation}\label{24MasterPKTS}
  \begin{aligned}
    \widetilde{\widehat{P}}(k,u,s)=\frac{1-\widehat{\omega}(u,s)}{s}+\frac{1-\widehat{\phi}(s)}{s} \frac{\widetilde{f}(k)\widehat{\omega}(u,s)}{1-\widetilde{f}(k) \widehat{\phi}(s)}.
  \end{aligned}
\end{equation}
Equation \eqref{24MasterPKTS} is a generalized Montroll Weiss equation describing ACTRW. The first term on the right-hand side of Eq.~\eqref{24MasterPKTS} corresponds to the case where the number of renewals is zero, describing non-moving particles. The inverse Fourier-Laplace transform is straightforward for this term. Below, we focus on the moving term in Eq.~\eqref{24MasterPKTS}. Unlike Eq.~\eqref{ffeq}, we consider the scaling when $x-at/\langle\tau\rangle$ is of the order of $t$.
Substituting Eq.~\eqref{firtwaitingtime} into the second term of Eq.~\eqref{24MasterPKTS}, we obtain
\begin{equation}\label{ddFLPxtta}
  \begin{split}
    \widetilde{\widehat{P}}(k,u,s)_{{\rm m}}&\sim\frac{\widehat{\phi}(s)-\widehat{\phi}(u)}{u-s} \frac{1-\widehat{\phi}(s)}{s(1-\widehat{\phi}(u))}
    \frac{\widetilde{f}(k)}{1-\widetilde{f}(k) \widehat{\phi}(s)}.
  \end{split}
\end{equation}
The subscript "m" denotes the moving term of Eq.~\eqref{24MasterPKTS}.
Recall that we are interested in the scaling when both $s$ and $k$ are small and comparable. Using Eq.~\eqref{LapPower}, we can express Eq.~\eqref{ddFLPxtta} as
\begin{equation}\label{needdetail_1}
  \begin{split}
    \widetilde{\widehat{P}}(k,u,s)_{{\rm m}}\sim& \frac{\widehat{h}(u)}{u-s}\left[\frac{\langle\tau\rangle b_{\alpha}s^{\alpha}}{\langle\tau\rangle s-ika}-\frac{ika\langle\tau\rangle b_{\alpha}s^{\alpha}}{(\langle\tau\rangle s-ika)^2}\right]\\
    &+\frac{1}{u-s}\frac{ika b_{\alpha}s^{\alpha-1}}{(\langle\tau\rangle s-ika)^2}
  \end{split},
\end{equation}
where $\widehat{h}(u) = 1/(1 - \widehat{\phi}(u))$. The derivation of Eq.~\eqref{needdetail_1} is provided in Appendix \ref{appA}.
Taking the inverse Laplace transform of Eq.~\eqref{needdetail_1} with respect to $u$, we obtain
%\begin{equation}
%  \begin{aligned}
%    \widetilde{\widehat{P}}(k,t_a,s)\sim&h(t_a)\ast\mathrm{exp\left(t_a s\right)}\frac{\langle\tau\rangle b_{\alpha}s^{\alpha}}{\langle\tau\rangle s-ika}\\
%    &-h(t_a)\ast\mathrm{exp\left(t_a s\right)}\frac{ika\langle\tau\rangle b_{\alpha}s^{\alpha}}{(\langle\tau\rangle s-ika)^2}\\
%    &+\mathrm{exp\left(t_a s\right)}\frac{ika b_{\alpha}s^{\alpha-1}}{(\langle\tau\rangle s-ika)^2}
%    \label{inverseTA}.
%  \end{aligned}
%\end{equation}
\begin{equation}\label{inverseTA}
  \begin{aligned}
    \widetilde{\widehat{P}}(k,t_a,s)\sim&\dfrac{\mathrm{exp}\left(t_a s\right)-1}{s}\left(\frac{b_{\alpha}s^{\alpha}}{\langle\tau\rangle s-ika}-\frac{ika b_{\alpha}s^{\alpha}}{(\langle\tau\rangle s-ika)^2}\right)\\
    &+\mathrm{exp}\left(t_a s\right)\frac{ika b_{\alpha}s^{\alpha-1}}{(\langle\tau\rangle s-ika)^2},
  \end{aligned}
\end{equation}
where we used the relation $\mathop{\mathcal{L}}\nolimits_{u\rightarrow t_a}^{-1}[1/(u-s)]=\exp(t_a s)$ and $\widehat{h}(u)\sim 1/(\langle\tau\rangle u)$. We assign the operators $\mathcal{L}^{-1}$ and $\mathcal{F}^{-1}$ as the inverse Laplace and the inverse Fourier transform, respectively.
Recall that the inverse Laplace transform of $\exp(t_a s) / (s - ik a / \langle\tau\rangle)^2$ results in $(t + t_a) \exp\left[ ika (t + t_a) / \langle\tau\rangle \right]$. Furthermore, the inverse Fourier transform of $\exp\left[ ika t / \langle\tau\rangle \right]$ corresponds to a delta function, and the inverse Fourier transform of the operator $ik \widetilde{f}(k)$ leads to the spatial derivative in $x$-space, i.e.,  $\mathcal{F}^{-1}[ik \widetilde{f}(k)]=df(x)/dx$. Thus, $P(x,t_a,t)$ decays as
%\begin{equation}
%  \begin{aligned}
%    P(x,t_a,t)\sim&\dfrac{\langle\tau\rangle b_{\alpha}h(t_a)}{a\Gamma(-\alpha)}\ast\dfrac{1}{(t+t_a-\langle\tau\rangle x/a)^{1+\alpha}}\\
%    &+\dfrac{b_{\alpha}h(t_a)}{\Gamma(-\alpha)}\ast\dfrac{\partial}{\partial x}\dfrac{\langle\tau\rangle x/a}{(t+t_a-\langle\tau\rangle x/a)^{1+\alpha}}\\
%    &-\dfrac{b_{\alpha}}{\langle\tau\rangle\Gamma(1-\alpha)}\dfrac{\partial}{\partial x}\dfrac{\langle\tau\rangle x/a}{(t+t_a-\langle\tau\rangle x/a)^{\alpha}}.\\
%    \label{endinverseX}
%  \end{aligned}
%\end{equation}
\begin{equation}
  \begin{aligned}    P(x,t_a,t)_{{\rm m}}\sim&\dfrac{\tau_0^{\alpha}}{a}\dfrac{1}{(t-\langle\tau\rangle \frac{x}{a})^{\alpha}}+\dfrac{\tau_0^{\alpha}}{a}\dfrac{d}{d x}\dfrac{x}{(t-\langle\tau\rangle \frac{x}{a})^{\alpha}}\\
    &-\dfrac{\tau_0^{\alpha}}{a}\dfrac{1}{(t+t_a-\langle\tau\rangle \frac{x}{a})^{\alpha}}.\\
  \end{aligned}
\end{equation}
By taking the derivatives with respect to $x$ and adding the non-moving term of Eq.~\eqref{24MasterPKTS}, there exists
%\begin{widetext}
%\begin{equation}
%\begin{aligned}
%P(x,t_a,t)\sim&h(t_a)\ast\dfrac{\langle\tau\rangle b_{\alpha}}{a\Gamma(-\alpha)}\left[\dfrac{2}{(t+t_a-\langle\tau\rangle \frac{x}{a})^{1+\alpha}}-\dfrac{(-1-\alpha)\langle\tau\rangle \frac{x}{a}}{(t+t_a-\langle\tau\rangle \frac{x}{a})^{2+\alpha}}\right]-\dfrac{b_{\alpha}\langle\tau\rangle}{\Gamma(1-\alpha)a}\\
%&\times\left[\dfrac{1}{(t+t_a-\langle\tau\rangle \frac{x}{a})^{\alpha}}+\dfrac{\alpha\langle\tau\rangle \frac{x}{a}}{(t+t_a-\langle\tau\rangle \frac{x}{a})^{1+\alpha}}\right]+h(t_a)\ast \dfrac{(\tau_0)^{\alpha}}{(t+t_a)^{\alpha}}\delta(x).
% \end{aligned}
%\end{equation}
%\end{widetext}
\begin{widetext}
\begin{equation}\label{endinverseX}
\begin{aligned}
P(x,t_a,t)\sim&\dfrac{\tau_0^{\alpha}}{a}\dfrac{1}{(t-\langle\tau\rangle x/a)^{\alpha}}+\dfrac{\tau_0^{\alpha}}{\langle\tau\rangle}\left[\dfrac{\langle\tau\rangle}{a}\dfrac{1}{(t-\langle\tau\rangle x/a)^{\alpha}}+\dfrac{\langle\tau\rangle^2 x\alpha}{a^2}\dfrac{1}{(t-\langle\tau\rangle x/a)^{1+\alpha}}\right]\\
&-\dfrac{\tau_0^{\alpha}}{a}\dfrac{1}{(t+t_a-\langle\tau\rangle x/a)^{\alpha}}+\dfrac{\tau_0^{\alpha}}{\langle\tau\rangle(1-\alpha)}\left[(t+t_a)^{1-\alpha}-t^{1-\alpha}\right]\delta(x).
 \end{aligned}
\end{equation}
\end{widetext}
From Eq.~\eqref{endinverseX}, we introduce the dimensionless random variable $\zeta = 1 - (x/a)/(t/\langle\tau\rangle)$, and obtain
%\begin{equation}\label{endPxtta1}
%  \begin{aligned}
%    &P(x,t_a,t)\sim \frac{\tau_0^{\alpha}}{at^{\alpha}}\zeta^{-1-\alpha}\left[\zeta+\alpha\Big(1-\zeta+\frac{t_a}{t}\Big)\right]\\
%    &+\dfrac{\alpha\tau_0^{\alpha}\langle\tau\rangle}{at^{1+\alpha}}h(t_a)\ast\zeta^{-2-\alpha}\left[2\zeta+(1+\alpha)\Big(1-\zeta+\frac{t_a}{t}\Big)\right]  \\
%    &+h(t_a)\ast \dfrac{\tau_0^{\alpha}}{(t+t_a)^{\alpha}}\delta(x).
%  \end{aligned}
%\end{equation}
\begin{equation}\label{endPxtta1}
  \begin{aligned}
    P(x,t_a,t)\sim& \dfrac{\tau_0^{\alpha}}{at^{\alpha}\zeta^{1+\alpha}}\Big[\alpha+(2-\alpha)\zeta\Big]-\frac{\tau_0^{\alpha}}{at^{\alpha}}\Big(\zeta+\dfrac{t_a}{t}\Big)^{-\alpha}\\
    &+\dfrac{\tau_0^{\alpha}}{\langle\tau\rangle(1-\alpha)}\left[(t+t_a)^{1-\alpha}-t^{1-\alpha}\right]\delta(x).
  \end{aligned}
\end{equation}
Equation~\eqref{endPxtta1} is valid for large $t$ and $t_a$, independent of the relationship between these two variables. When $t_a=0$, it reduces to the theoretical prediction for the CTRW model, as discussed in \cite{Wang2019Transport}, with the delta function disappearing. It can be seen that the above equation is non-integrable concerning $\zeta$, in that sense, Eq.~\eqref{endPxtta1} is called an infinite density \cite{Rebenshtok2014Infinite}.
As shown in Fig.~\ref{RareEVENTSofX}, the far tail of the positional distribution becomes sensitive to the aging time of the systems and is well described by Eq.~\eqref{endPxtta1}.
This is because the MSD is highly influenced by the slow-moving particles that lag significantly behind the mean position.

%%111111111111111111111111111111111

%%111111111111111111111111
We now examine two limiting laws of Eq.~\eqref{endPxtta1} to highlight the relationship between the CTRW model and the equilibrium CTRW model, specifically for $t_a \ll t$ and $t \ll t_a$, which correspond to weakly aging and strongly aging systems, respectively.

\begin{figure}[htp]
  \centerline{\includegraphics[width=22pc]{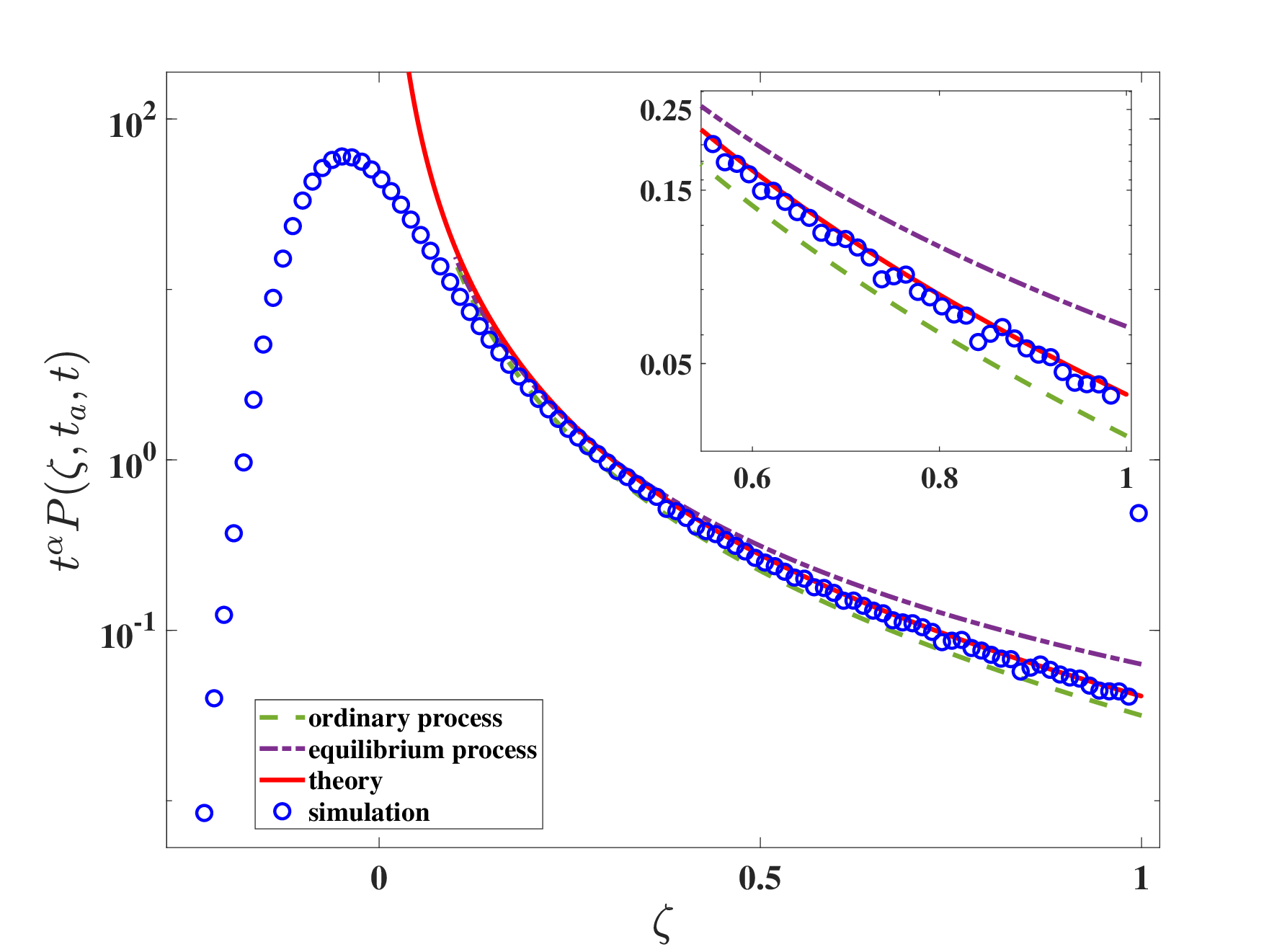}}
  \caption{The Scaled PDF when $t_a\ll t$, where $\zeta=1-(x/a)/(t/\langle\tau\rangle).$  The solid line, calculated from Eq.~\eqref{Pxtta_sllu}, represents the behavior of $x$
 for weakly aging systems (see inset). The other two lines, derived from the first term of Eq.~\eqref{Pxtta_sllu} and the first term of Eq.~\eqref{Pxtta_ulls}, correspond to rare events for ordinary and equilibrium processes, respectively. The symbols indicate simulation results obtained from $10^7$ realizations. The parameters used are $\alpha=3/2$, $a=1$, $\sigma=1$, $t_a=200$, $t=1000$, and $\tau_0=0.1$. }
  \label{WeakAging}
\end{figure}

The key idea, based on Eq.~\eqref{needdetail_1}, is to calculate the following term
\begin{equation}\label{convolution}
  \dfrac{\widehat{h}(u)}{u-s}=\dfrac{1}{u-s}\dfrac{1}{1-\widehat{\phi}(u)}\sim\dfrac{1}{u-s}\dfrac{1}{\langle\tau\rangle u-b_{\alpha}u^{\alpha}}.
\end{equation}
When both $s$ and $u$ are small but $s \ll u $, in real space, it corresponds to $ 0 \ll t_a \ll t $. The asymptotic behavior of Eq.~\eqref{convolution} in this regime is given by
\begin{equation}\label{sllu}
  \dfrac{\widehat{h}(u)}{u-s}=\dfrac{1}{u}\Big(1+\dfrac{s}{u}\Big)\Big(\dfrac{1}{\langle\tau\rangle u}+\dfrac{b_{\alpha}}{\langle\tau\rangle^2}u^{\alpha-2}\Big).
\end{equation}
Substituting Eq.~\eqref{sllu} into Eq.~\eqref{needdetail_1} and performing the inverse Fourier-Laplace transforms yield
\begin{equation}\label{Pxtta_sllu}
  P(x,t_a,t)_{{\rm m}}\sim \frac{\tau_0^{\alpha}}{at^{\alpha}}\mathop{\mathcal{L}}\nolimits_{{\rm or},\alpha}(\zeta)+t_a\frac{\alpha\tau_0^{\alpha}}{at^{1+\alpha}}\zeta^{-1-\alpha}
\end{equation}
with \begin{equation}\label{or}
  \mathop{\mathcal{L}}\nolimits_{{\rm or},\alpha}(\zeta)=\alpha\zeta^{-1-\alpha}+(1-\alpha)\zeta^{-\alpha}
\end{equation}
and $\zeta = 1 - (x/a)/(t/\langle\tau\rangle)$.
For detailed calculations, refer to Appendix \ref{APPC}. The first term on the right-hand side of Eq.~\eqref{Pxtta_sllu} corresponds to rare fluctuations for the CTRW model as discussed in \cite{Wang2019Transport}, and the second term becomes negligible as $t_a \to 0$, which is as anticipated. The comparison between Eq.~\eqref{Pxtta_sllu} and simulations is illustrated in Fig.~\ref{WeakAging}, showing an excellent agreement.

Here, we consider the opposite scenario where both $s$ and $u$ are small, but $u \ll s$. By using the approximation $ 1/(1-y) \approx 1 + y $ for $ y \to 0 $, we obtain
\begin{equation}\label{ulls}
  \dfrac{\widehat{h}(u)}{u-s}\sim -\dfrac{1}{s}\Big(1+\dfrac{u}{s}\Big)\Big(\dfrac{1}{\langle\tau\rangle u}+\dfrac{b_{\alpha}}{\langle\tau\rangle^2}u^{\alpha-2}\Big).
\end{equation}
Based on Eq. \eqref{ulls}, Eq. \eqref{needdetail_1} leads to
\begin{equation}\label{Pxtta_ulls}
  P(x,t_a,t)_{\rm{m}}\sim \dfrac{\tau_0^{\alpha}}{a t^{\alpha}}\mathop{\mathcal{L}}\nolimits_{{\rm eq},\alpha}(\zeta)-\dfrac{\tau_0^{2\alpha}\mathop{\mathcal{H}}(\zeta)}{a\langle\tau\rangle t^{\alpha-1}t_a^{\alpha}(1-\alpha)}
\end{equation}
with
\begin{equation}\label{eq}
  \mathop{\mathcal{L}}\nolimits_{{\rm eq},\alpha}(\zeta)=\alpha\zeta^{-1-\alpha}+(2-\alpha)\zeta^{-\alpha}
\end{equation}
and
\begin{equation}
  \mathop{\mathcal{H}}(\zeta)=(3-\alpha)\zeta^{1-\alpha}-(1-\alpha)\zeta^{-\alpha}.
\end{equation}
The second term in Eq.~\eqref{Pxtta_ulls} is determined by the aging time $t_a$ and the observation time $t$. Taking $t_a \to \infty$, Eq.~\eqref{Pxtta_ulls} aligns with the theory for the equilibrium CTRW model discussed in \cite{Wang2019Transport}, since the second term disappears.

Using the far tail of the position provided in Eq.~\eqref{endinverseX}, the MSD of the position follows
\begin{equation}\label{MSDWEAK}
\begin{aligned}
    \langle(x(t)-\langle x(t)\rangle)^2\sim&\dfrac{2a^2 b_{\alpha}}{\langle\tau\rangle^3\Gamma(4-\alpha)}\left[t^{3-\alpha}+(t+t_a)^{3-\alpha}\right.\\
    &-\left.t_a^{3-\alpha}-(3-\alpha)(t+t_a)^{2-\alpha}t\right].
\end{aligned}
\end{equation}
For further details, refer to Appendix \ref{appD}. Similar to the discussion on rare events in the CTRW, the MSD of the ACTRW is also governed by these slowly moving particles near the initial position. It is worth noting that Eq.~\eqref{MSDWEAK} can also be derived using the relationship between the moments and the characteristic function.

\section{Strong relation between rare events of the position and the number of renewals}\label{Strongrelation}
In this section, we focus on the far tail of renewals, highlighting the strong relationship between the position and the number of renewals. The scaling considered here is when $ N - t/\langle\tau\rangle $ is of the order of the observation time $t$.

Starting from Eq.~\eqref{QusMving} and applying the Laplace transform with respect to $N$, we obtain
\begin{equation}\label{24sect5001}
  \begin{aligned}
    \widehat{Q}_{t_a,s}(v)\sim&\widehat{\omega}(t_a,s)\frac{1-\widehat{\phi}(s)}{s}\\
    &\times\int_{0}^{\infty} \exp \{-vN+N\mathrm{ln}[\widehat{\phi}(s)]\} \mathrm{d}N.
  \end{aligned}
\end{equation}
Given our interest in the long-time behavior, where
$N$ can be treated as a continuous variable, performing the integral over
$N$ in Eq.~\eqref{24sect5001} yields
\begin{equation}\label{24sect5002}
  \begin{aligned}
    \widehat{Q}_{t_a,s}(v)\sim&\widehat{\omega}(t_a,s)\frac{1-\widehat{\phi}(s)}{s}\frac{1}{v-\ln(\widehat{\phi}(s))}.
  \end{aligned}
\end{equation}
To further analyze the system, we introduce the new random variable $\epsilon = N - t/\langle\tau\rangle$ and investigate its statistics. After performing the Fourier transform with respect to $\epsilon$ (i.e., $\epsilon \rightarrow k$) and the double Laplace transform with respect to $t$ and $t_a$ (i.e., $t \rightarrow s$, $t_a \rightarrow u$), the PDF of $\epsilon$, derived from Eq.~\eqref{24sect5002} in Fourier-Laplace spaces, is given by

\begin{equation}\label{24sect5003}
  \begin{aligned}
    \widehat{Q}_{u,s}(k)\sim &\frac{\widehat{\phi}(s+\frac{ik}{\langle\tau\rangle})-\widehat{\phi}(u)}{(u-s-\frac{ik}{\langle\tau\rangle})(1-\widehat{\phi}(u))}\frac{1-\widehat{\phi}(s+\frac{ik}{\langle\tau\rangle})}{s+\frac{ik}{\langle\tau\rangle}}\\
    &\times\frac{1}{-ik-\mathrm{ln}(\widehat{\phi}(s+\frac{ik}{\langle\tau\rangle}))}.
  \end{aligned}
\end{equation}
Mathematically, the solution of $ Q_{t_a,t}(N) $ can be approximated by Eq.~\eqref{QtatC}, with $Q_{t}(N) $ following the asymmetric L{\'e}vy stable law. Additionally, in the long time limit, the central part of the PDF for the number of renewals is not significantly affected by the system's aging time, resulting in $ Q_{t_a,t}(N) \sim \mathcal{L}_{\alpha}\left((N - t/\langle\tau\rangle)/(t/\bar{t})^{1/\alpha}\right)/(t/\bar{t})^{1/\alpha} $, being the same as in Eq.~\eqref{QtN}. However, our focus is not on the central part but rather the far tail.

\begin{figure}[!ht]
  \centerline{\includegraphics[width=22pc]{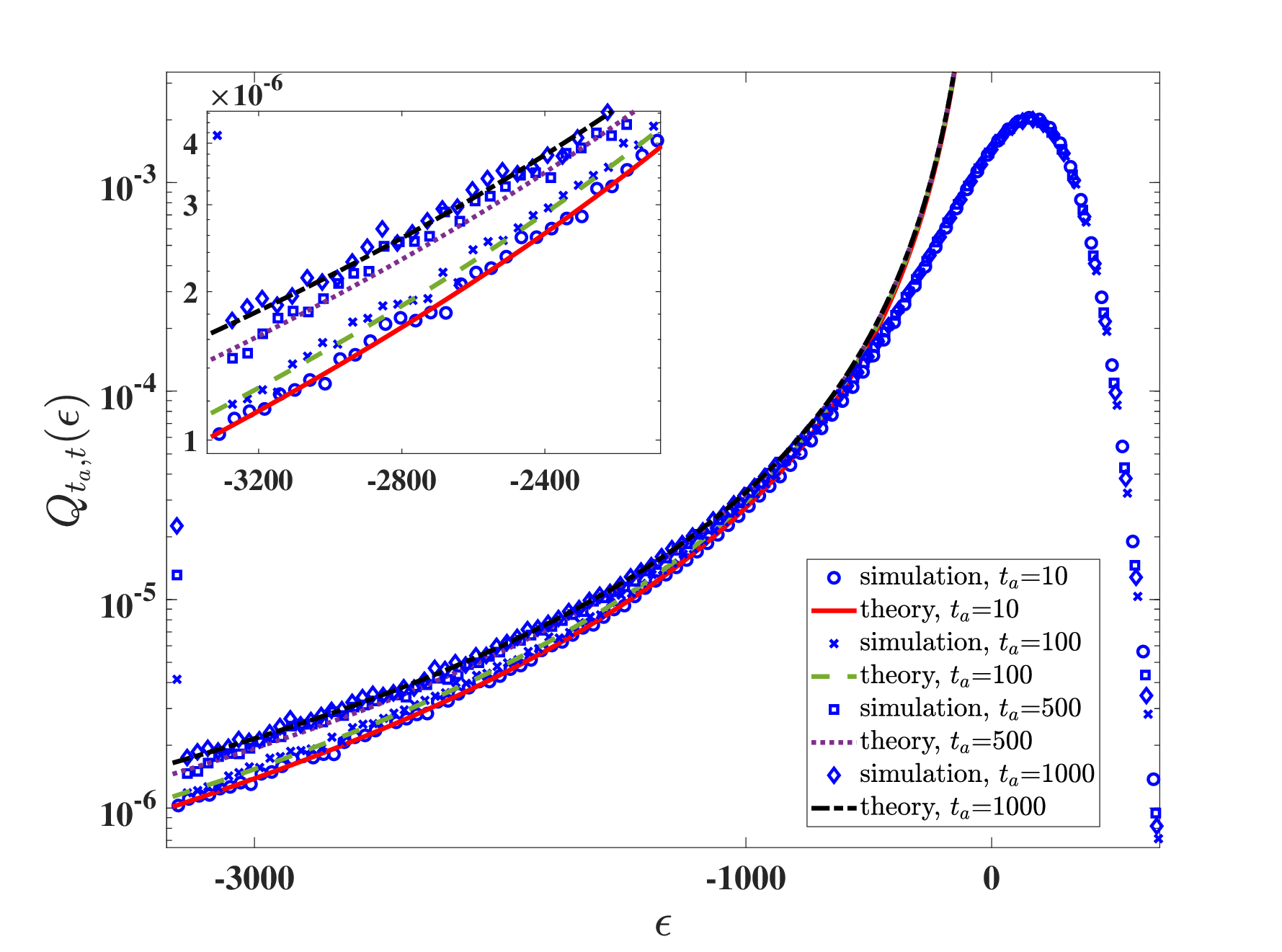}}
  \caption{The plot of $Q_{t_a,t}(\epsilon)$ for various $t_a$ with $\epsilon=N-t/\langle\tau\rangle$. The theories plotted by lines are obtained from Eq. \eqref{convolutionPNT}, and symbols illustrate the corresponding simulations. Here $\alpha=3/2$, $t=1000$, $\tau_0=0.1$ and $t_a=10, 100, 500$, and $1000$.}
  \label{Different_ta_N}
\end{figure}

Using the relation $\widehat{\phi}(s+ik/\langle\tau\rangle)-\widehat{\phi}(u)=\widehat{\phi}(s+ik/\langle\tau\rangle)-1+1-\widehat{\phi}(u)$, and assuming that $u$ and $s$ are small and comparable, Eq.~\eqref{24sect5003} can be divided into two terms
\begin{equation}\label{Q}
  \begin{aligned}
    \widehat{Q}_{u,s}(k)\sim&\frac{b_{\alpha}\langle\tau\rangle^{-1}}{u-s-\frac{ik}{\langle\tau\rangle}}\left[\frac{\Big(s+\frac{ik}{\langle\tau\rangle}\Big)^{\alpha}}{s^2}-\frac{\Big(s+\frac{ik}{\langle\tau\rangle}\Big)^{\alpha-1}}{s}\right]\\
    +&\frac{\widehat{h}(u)b_{\alpha}}{u-s-\frac{ik}{\langle\tau\rangle}}\left[\frac{\Big(s+\frac{ik}{\langle\tau\rangle}\Big)^{\alpha+1}}{s^2}-\frac{2\Big(s+\frac{ik}{\langle\tau\rangle}\Big)^{\alpha}}{s}\right].
  \end{aligned}
\end{equation}
Recall that when $t_a$ is long, $\widehat{h}(u)$ provided in the above equation can be approximated as $\widehat{h}(u) = 1/(1-\widehat{\phi}(u)) \sim 1/(\langle\tau\rangle u)$. Then, by performing the inverse Laplace transform with respect to $u$, the inverse Fourier transform with respect to $k$, the inverse Laplace transform with respect to $s$ (i.e., $u \rightarrow t_a$, $k \rightarrow \epsilon$, and $s \rightarrow t$), we obtain the main result of this section
\begin{equation}\label{convolutionPNT}
  \begin{aligned}
    Q_{t_a,t}(\epsilon)\sim&\tau_0^{\alpha}\left[\frac{(2-\alpha)}{(-\langle\tau\rangle\epsilon)^{\alpha}}+\frac{\alpha t}{(-\langle\tau\rangle\epsilon)^{1+\alpha}}\right]-\frac{\tau_0^{\alpha}}{(t_a-\langle\tau\rangle\epsilon)^{\alpha}}.\\
  \end{aligned}
\end{equation}
%with $\zeta=t_a-\langle\tau\rangle\epsilon$.
    Eq.~\eqref{convolutionPNT} is valid when both $t_a$ and $t$ are large. For $\epsilon<0$, i.e., $N < t/\langle \tau \rangle$, Eq.~\eqref{convolutionPNT} effectively describes the far tail of the number of renewals rather than the central part, as plotted in Fig.~\ref{Different_ta_N}. Notably, $N$ is expected to exhibit a cutoff at the tail $N=0$. It can be seen that the strong relation between the position and the number of renewals is $x\sim aN\sim a t/\langle\tau\rangle$ obtained using large deviations.

Similarly to the discussion of the position, we further explore the limits of Eq.~\eqref{convolutionPNT}, specifically for the cases $t_a \ll t$ and $t\ll t_a$. These two limits arise from how we consider the relationship between $u$ and $s$ in Eq.~\eqref{Q}. To be more exact, how to approximate the term $\widehat{h}(u)/(u-s-ik/\langle\tau\rangle).$

\begin{figure}[!ht]
  \centerline{\includegraphics[width=22pc]{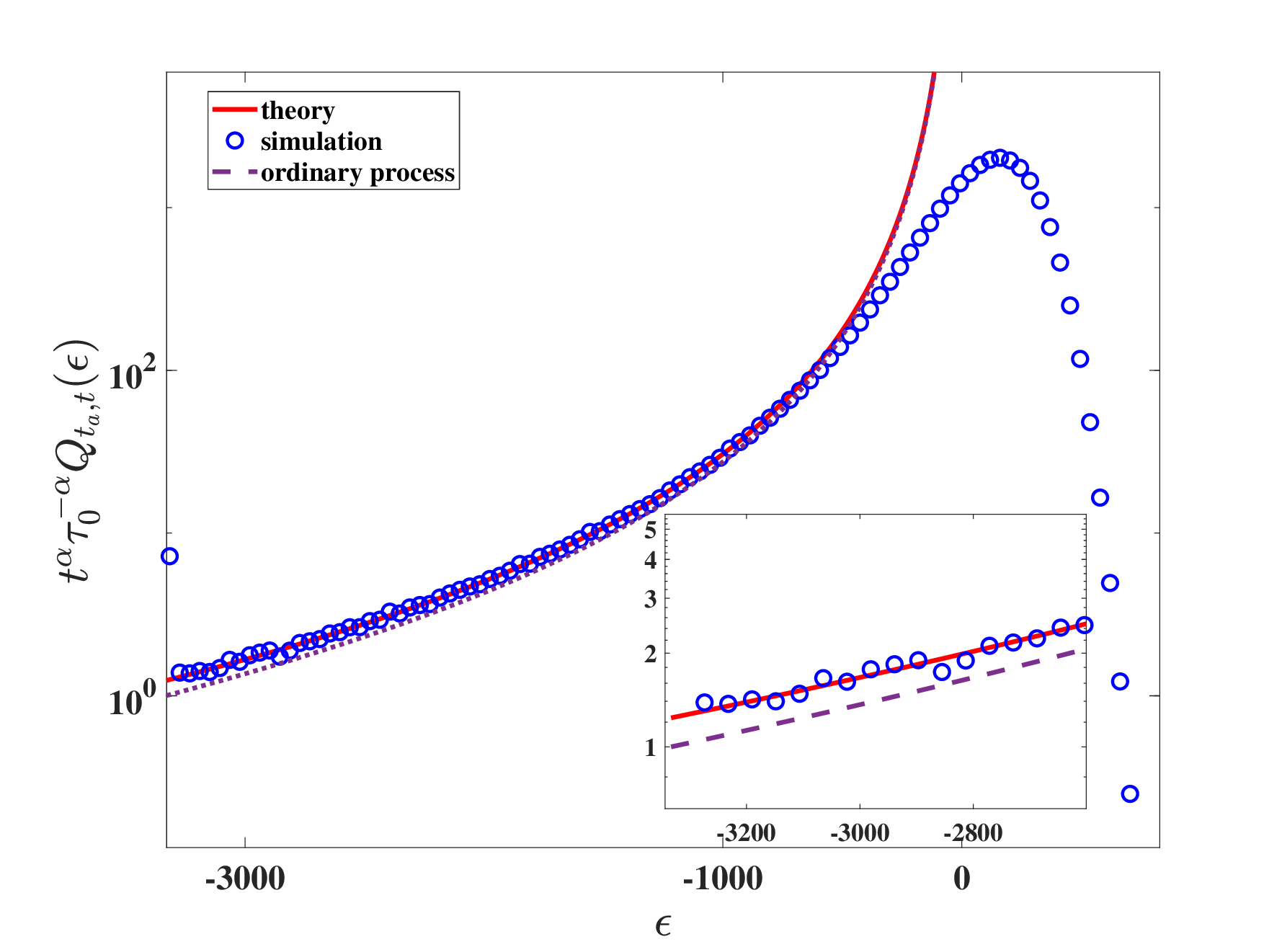}}
  \caption{The behavior of $Q_{t_a,t}(\epsilon)$ for $t_a \ll t$ with $\epsilon = N - t/\langle \tau \rangle$. The solid line represents the theoretical prediction calculated from Eq.~\eqref{PNTtallt}, while the dashed line corresponds to the description of rare events for a normal renewal process predicted by Eq.~\eqref{orPNT}. The parameters used are $\alpha = 3/2$, $t = 1000$, $\tau_0 = 0.1$, and $t_a = 200$.}
  \label{WeakAgingN}
\end{figure}

When $t_a\ll t$, in Laplace space, it corresponds to $s\ll u$. For this case, we have the following relation
\begin{equation}\label{talltN}
  \begin{aligned}
    \frac{\widehat{h}(u)}{u-s-\frac{ik}{\langle\tau\rangle}}\sim \left(1+\frac{s+\frac{ik}{\langle\tau\rangle}}{u}\right)\left(\frac{1}{\langle\tau\rangle u^2}+\frac{b_{\alpha}u^{\alpha-3}}{\langle\tau\rangle^2}\right).
  \end{aligned}
\end{equation}
Utilizing Eqs.~\eqref{Q} and \eqref{talltN}, we obtain the  far tail of the $\epsilon$
\begin{equation}\label{PNTtallt}
    Q_{t_a,t}(\epsilon)\sim\frac{\tau_0^{\alpha}\alpha}{(-\langle\tau\rangle\epsilon)^{1+\alpha}}\left[\dfrac{1-\alpha}{\alpha}(-\langle\tau\rangle\epsilon)+t+t_a\right].
\end{equation}
See this equation plot in Fig.~\ref{WeakAgingN}. In the particular case of $t_a=0$, Eq.~\eqref{PNTtallt} reduces to the case of the normal renewal process
\begin{equation}\label{orPNT}
    Q_{t}(\epsilon)\sim\tau_0^{\alpha}\left[(1-\alpha)(-\langle\tau\rangle\epsilon)^{-\alpha}+\alpha t(-\langle\tau\rangle\epsilon)^{-1-\alpha}\right]
\end{equation}
as discussed in \cite{Wang2018Renewal}.

\begin{figure}[!ht]
  \centerline{\includegraphics[width=22pc]{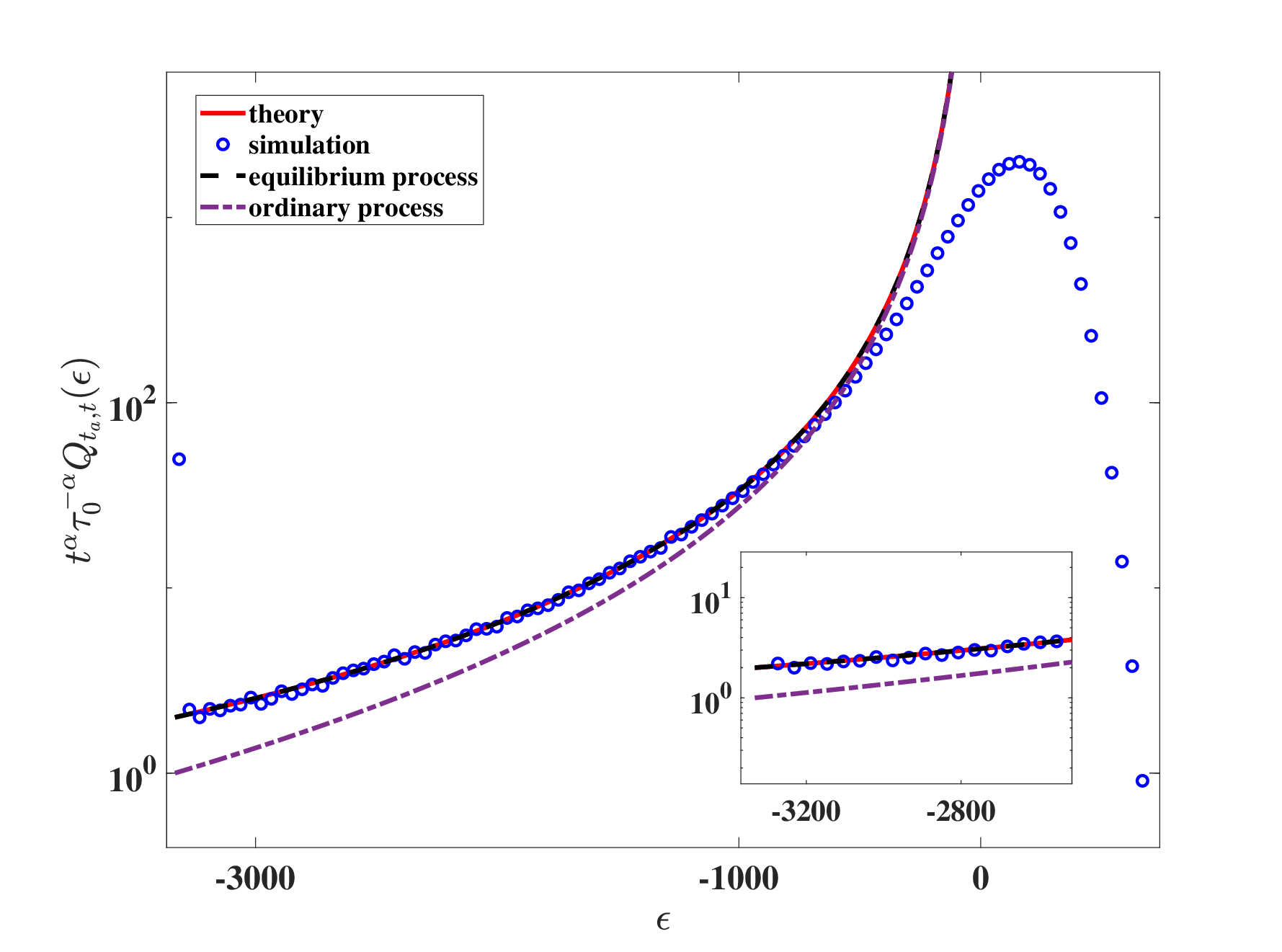}}
  \caption{The behavior of $Q_{t_a,t}(\epsilon)$ for $t_a \gg t$ with $\epsilon = N - t/\langle \tau \rangle$. The solid line represents the analytical prediction calculated from Eq.~\eqref{PNTtllta}, while the dashed line, corresponding to the equilibrium process, is obtained from Eq.~\eqref{PNTnota}. The dashed-dotted line depicts the theory for the normal renewal process given by Eq.~\eqref{orPNT}. Symbols indicate simulation results based on $3 \times 10^6$ realizations.
  Here we choose $\alpha=3/2$, $t_a=10000$, $t=1000,$ and $\tau_0=0.1.$
 }
  \label{StorngAgingN}
\end{figure}

Next, we consider the second situation, which assumes $u \ll s$ and both $s$ and $u$ are small. Similarly, we get
\begin{equation}
  \begin{aligned}
    &\frac{\widehat{h}(u)}{u-s-\frac{ik}{\langle\tau\rangle}}\\
    &\sim -\left(\frac{1}{s+\frac{ik}{\langle\tau\rangle}}+\frac{u}{(s+\frac{ik}{\langle\tau\rangle})^2}\right)\left(\frac{1}{\langle\tau\rangle u}+\frac{b_{\alpha}u^{\alpha-2}}{\langle\tau\rangle}\right).
  \end{aligned}
\end{equation}
Thus, the far tail follows
\begin{equation}\label{PNTtllta}
  \begin{aligned}
    Q_{t_a,t}(\epsilon)
    \sim&\frac{\tau_0^{\alpha}\alpha}{(-\langle\tau\rangle\epsilon)^{1+\alpha}}\left[\dfrac{2-\alpha}{\alpha}(-\langle\tau\rangle\epsilon)+t\right]\\
    &+\dfrac{\tau_0^{2\alpha}}{t_a^{\alpha}\langle\tau\rangle(-\langle\tau\rangle\epsilon)^{\alpha}}\left[\dfrac{3-\alpha}{\alpha-1}(-\langle\tau\rangle\epsilon)+t\right].
  \end{aligned}
\end{equation}
See the plot in Fig.~\ref{StorngAgingN}. In the particular case of $t_a\to\infty$, Eq.~\eqref{PNTtllta} leads to
\begin{equation}\label{PNTnota}
  \begin{aligned}
    Q_{\infty,t}(\epsilon)
    \sim&\frac{\tau_0^{\alpha}\alpha}{(-\langle\tau\rangle\epsilon)^{1+\alpha}}\left[\dfrac{2-\alpha}{\alpha}(-\langle\tau\rangle\epsilon)+t\right],
  \end{aligned}
\end{equation}
describing the rare fluctuations of $N$ in the context of the equilibrium renewal process.

\section{Conclusion}\label{Conclusion}
In this manuscript, we focus on the ACTRW in one dimension, analyzing the statistical properties of the position and the number of renewals. We demonstrate that when waiting times have a finite mean, but an infinite variance, and displacements follow a Gaussian distribution with a non-zero mean, the behavior of a particle diffusing in a disordered system with traps can be described by the aging fractional advection-diffusion equation \eqref{ffeq}. Notably, the fractional operator in this context is spatial rather than temporal, indicating that a power-law type distribution of displacements is not a prerequisite for the spatial operator, as discussed in \cite{Wang2020Fractional,Wang2024Fractional}. The fractional equation presented here can be generalized further: it can accommodate the case of displacements with a narrow distribution and non-zero mean, and it allows for modification to the statistics over the time interval $(-t_a, 0)$, as explored in Fig.~\ref{ComplexAgingFokker}.  The diffusion equation for random walks can also be derived from the generalized master equation \cite{Metzler2000Generalized,Allegrini2003Generalized}. For instance, when waiting times have an infinite mean, the generalized master equation serves as a framework that unifies fractional calculus with CTRW. However, the approach in this manuscript differs; we do not derive the diffusion equation directly from the master equation. Instead, using the subordination method, we show that a fractional space diffusion equation, commonly employed to describe the statistics of L{\'e}vy flights, can also effectively characterize the ACTRW model.

The manuscript demonstrates that typical fluctuations are largely insensitive to the initial conditions of the process, but rare fluctuations are significantly affected by them. Consequently, the theory presented here provides a method to distinguish between different aging times. This conclusion applies to both the positional distribution in Eq.~\eqref{endPxtta1} and the number of renewals in Eq.~\eqref{convolutionPNT}. As expected, our results align with those in \cite{Wang2019Transport} when considering the limits $t_a \to 0$ or $t_a \to \infty$. It is important to note that the rare events of renewals discussed in this manuscript differ from those in \cite{Barkai2020Packets,Wang2024Statistics}, which focus on the short-time limit statistics.

\section*{Acknowledgments}
The work is supported by the National Natural Science Foundation of China under Grant No.
12105243 and the Zhejiang Province Natural Science Foundation LQ$22$A$050002$.

\appendix
%\onecolumngrid
\begin{appendices}

\section{The derivation of Eq. \texorpdfstring{\eqref{needdetail_1}}{}}\label{appA}
We provide a detailed derivation of Eq.~\eqref{needdetail_1} in the main text. Starting from Eq.~\eqref{ddFLPxtta}, we decompose the expression into two parts by rewriting it as $(\widehat{\phi}(s) - \widehat{\phi}(u))/(u - s) = (\widehat{\phi}(s) - 1 + 1 - \widehat{\phi}(u))/(u - s)$, which allows us to proceed with the following form
\begin{equation}
\widetilde{\widehat{P}}_{\rm m}(k,u,s)=\widetilde{\widehat{P}}_{\rm m1}(k,u,s)+\widetilde{\widehat{P}}_{\rm m2}(k,u,s)
\end{equation}
with
\begin{equation}
  \begin{aligned}
    \widetilde{\widehat{P}}_{\rm m1}(k,u,s)=\frac{-[1-\widehat{\phi}(s)]^2}{s(u-s)(1-\widehat{\phi}(u))}\frac{\widetilde{f}(k)}{1-\widetilde{f}(k)\widehat{\phi}(s)}\label{P1}
  \end{aligned}
\end{equation}
and
\begin{equation}
  \begin{aligned}
    \widetilde{\widehat{P}}_{\rm m2}(k,u,s)=\frac{1-\widehat{\phi}(s)}{s}\frac{1}{u-s}\frac{\widetilde{f}(k)}{1-\widetilde{f}(k)\widehat{\phi}(s)}.\label{P2}
  \end{aligned}
\end{equation}
Inserting Eqs. \eqref{LapPower} and Eq. \eqref{GaussianF} into Eq. \eqref{P1}, there exists
\begin{equation}\label{P11}
  \begin{aligned}
\widetilde{\widehat{P}}_{\rm m1}(k,u,s)\sim\frac{\frac{\widehat{h}(u)}{u-s}(2\langle\tau\rangle b_{\alpha}s^{\alpha}-b_{\alpha}^{2}s^{2\alpha-1}-\langle\tau\rangle^{2}s)}{(\langle\tau\rangle s-ika)(1-\frac{b_{\alpha}s^{\alpha}}{\langle\tau\rangle s-ika})},
  \end{aligned}
\end{equation}
where we ignored the term $k^2$ since we
focus on the case where $\langle\tau\rangle s$ and $ika$ are comparable and small.
Note that  $s^{2\alpha-1}\ll s^{\alpha}$, and the same goes for other items that are $s^{3\alpha-1}/(\langle\tau\rangle s-ika)\varpropto s^{3\alpha-2}$ and $s^{2\alpha}/(\langle\tau\rangle s-ika)\varpropto s^{2\alpha-1}$. Besides, $s^{\alpha}\ll s$ and $(b_{\alpha}s^{\alpha})/(\langle\tau\rangle s-ika)\rightarrow 0$, hence, it could use $1/(1-y)\simeq 1+y$ with $y\rightarrow 0$ for the above equation. Eq. \eqref{P11} reduces to
\begin{equation}
  \begin{aligned}
    \widetilde{\widehat{P}}_{\rm m1}(k,u,s)&\sim \frac{\widehat{h}(u)}{(u-s)(s-\frac{ika}{\langle\tau\rangle})}\Big(2b_{\alpha}s^{\alpha}-\frac{\langle\tau\rangle b_{\alpha}s^{\alpha+1}}{\langle\tau\rangle s-ika}\Big).
  \end{aligned}
\end{equation}
Rewriting the above equation, we obtain
\begin{equation}
  \begin{aligned}
    \widetilde{\widehat{P}}_{\rm m1}(k,u,s)&\sim \frac{\widehat{h}(u)}{u-s}\left[\frac{\langle\tau\rangle b_{\alpha}s^{\alpha}}{\langle\tau\rangle s-ika}-\frac{ika\langle\tau\rangle b_{\alpha}s^{\alpha}}{(\langle\tau\rangle s-ika)^2}\right]\label{P12}.
  \end{aligned}
\end{equation}
Now, let's consider the second part, $\widetilde{\widehat{P}}_{\rm m2}(k, u, s)$. Following a similar approach as in Eq.~\eqref{P12}, we substitute Eqs.\eqref{LapPower} and \eqref{GaussianF} into Eq.~\eqref{P2}, yielding
\begin{equation}
  \begin{aligned}
    \widetilde{\widehat{P}}_{m2}(k,u,s)&\sim\frac{1}{u-s}\frac{ika b_{\alpha}s^{\alpha-1}}{(\langle\tau\rangle s-ika)^2}.\label{P22}
  \end{aligned}
\end{equation}
Based on the above equations, in Fourier-Laplace space, we obtain the equation describing rare fluctuation of the position
\begin{equation}
  \begin{aligned}
    \widetilde{\widehat{P}}_{\rm m}(k,u,s)&\sim\frac{\widehat{h}(u)}{u-s}\left[\frac{\langle\tau\rangle b_{\alpha}s^{\alpha}}{\langle\tau\rangle s-ika}-\frac{ika\langle\tau\rangle b_{\alpha}s^{\alpha}}{(\langle\tau\rangle s-ika)^2}\right]\\
    &+\frac{1}{u-s}\frac{ika b_{\alpha}s^{\alpha-1}}{(\langle\tau\rangle s-ika)^2},
  \end{aligned}
\end{equation}
which gives Eq. \eqref{needdetail_1} in main text.

\section{Details of calculations of Eqs. \texorpdfstring{\eqref{Pxtta_sllu}}{} and Eq. \texorpdfstring{\eqref{Pxtta_ulls}}{}}\label{APPC}
We begin by investigating the case where
$s\ll u$, and we derive the following expression from Eq. \eqref{needdetail_1}
\begin{equation}
\begin{aligned}
    	\widetilde{\widehat{P}}(k,u,s)_{\rm m}&\sim\dfrac{1}{\langle\tau\rangle u^2}(1+\dfrac{s}{u})(1+\dfrac{b_{\alpha}}{\langle\tau\rangle}u^{\alpha-1})\\
     &\times\left[\dfrac{\langle\tau\rangle b_{\alpha}s^{\alpha}}{\langle\tau\rangle s-ika}\right.-\left.\dfrac{ika\langle\tau\rangle b_{\alpha}s^{\alpha}}{(\langle\tau\rangle s-ika)^2}\right]\\
     &+(\frac{1}{u}+\frac{s}{u^2})\left[\frac{ika\langle\tau\rangle b_{\alpha}s^{\alpha-1}}{(\langle\tau\rangle s-ika)^2}\right].
\end{aligned}
\end{equation}
When $t$ and $t_a$ is large, i.e., $s,u\to 0$, we have
\begin{equation}\label{PKSus}
\begin{aligned}
    \widetilde{\widehat{P}}(k,u,s)_{\rm m}&\sim\dfrac{1}{\langle\tau\rangle u^2}\left[\dfrac{\langle\tau\rangle b_{\alpha}s^{\alpha}}{\langle\tau\rangle s-ika}-\dfrac{ika\langle\tau\rangle b_{\alpha}s^{\alpha}}{(\langle\tau\rangle s-ika)^2}\right]\\
    &+\frac{1}{u}(1+\frac{s}{u})\left[\frac{ika\langle\tau\rangle b_{\alpha}s^{\alpha-1}}{(\langle\tau\rangle s-ika)^2}\right].
\end{aligned}
\end{equation}
Utilizing the relation%here need modify
\begin{equation}
\mathop{\mathcal{F}}\nolimits_{k\rightarrow x}^{-1}\mathop{\mathcal{L}}\nolimits_{s\rightarrow t}^{-1}\left[\dfrac{s^{\alpha}}{s-\frac{ika}{\langle\tau\rangle}}\right]\sim \dfrac{\langle\tau\rangle}{a\Gamma(-\alpha)}\left(t-\frac{\langle\tau\rangle}{a}\right)^{-\alpha-1}
\end{equation}
and
\begin{equation}
    \mathop{\mathcal{F}}\nolimits_{k\rightarrow x}^{-1}\mathop{\mathcal{L}}\nolimits_{s\rightarrow t}^{-1}\left[\dfrac{-ika s^{\alpha}}{(s-\frac{ika}{\langle\tau\rangle})^2}\right]\sim \dfrac{1}{\Gamma(-\alpha)}\dfrac{d}{d x}\dfrac{\frac{\langle\tau\rangle}{a}x}{(t-\frac{\langle\tau\rangle}{a})^{\alpha+1}},
\end{equation}
%&\dfrac{\tau_0^{\alpha}}{at^{\alpha}}\left[(1-\alpha)\xi^{-\alpha}+\alpha\xi^{-\alpha-1}\right]\\
%&+\dfrac{\alpha \tau_0^{\alpha}t_a}{at^{\alpha+1}}\left[(1-\alpha)\xi^{-\alpha-1}+(1+\alpha)\xi^{-\alpha-2}\right]\\
%&-\dfrac{\alpha \tau_0^{\alpha}t_a}{at^{\alpha+1}}\left[(-\alpha)\xi^{-\alpha-1}+(1+\alpha)\xi^{-\alpha-2}\right]\\
%=&
where $\mathcal{F}^{-1}$ is Fourier transform and $\mathcal{L}^{-1}$ is Laplace transform, the inverse transform of Eq.~\eqref{PKSus} yields
\begin{equation}
	\begin{aligned}
    	P(x,t_a,t)_{m}\sim\dfrac{\tau_0^{\alpha}}{at^{\alpha}}\left[\frac{(1-\alpha)}{\zeta^{\alpha}}+\frac{\alpha}{\zeta^{\alpha+1}}\right]+\dfrac{\alpha \tau_0^{\alpha}t_a}{a(t\zeta)^{\alpha+1}}.
	\end{aligned}
\end{equation}
The above equation leads to Eq.~\eqref{Pxtta_sllu} in the main text.

Below, we consider another situation, i.e., $u\ll s$. Again from Eq. \eqref{needdetail_1}, we get
\begin{equation}
\begin{aligned}
    \widetilde{\widehat{P}}(k,u,s)_{\rm m}&\sim H(u,s)\left[\dfrac{ika\langle\tau\rangle b_{\alpha}s^{\alpha}}{(\langle\tau\rangle s-ika)^2}-\dfrac{\langle\tau\rangle b_{\alpha}s^{\alpha}}{\langle\tau\rangle s-ika}\right]\\
    &-\frac{1}{s}\left(1+\frac{u}{s}\right)\left[\frac{ika\langle\tau\rangle b_{\alpha}s^{\alpha-1}}{(\langle\tau\rangle s-ika)^2}\right],
\end{aligned}
\end{equation}
with $H(u,s)=\frac{1}{s\langle\tau\rangle u}(1+\frac{u}{s})(1+\frac{b_{\alpha}}{\langle\tau\rangle}u^{\alpha-1})$.
Applying the inverse transform and rearranging it,  we have
\begin{equation}
\begin{aligned}
    P(x,t_a,t)_{\rm m}\sim&\dfrac{\tau_0^{\alpha}}{at^{\alpha}}\left[(2-\alpha)\zeta^{-\alpha}+\alpha\zeta^{-\alpha-1}\right]\\
    -&\dfrac{\tau_0^{2\alpha}\left[(3-\alpha)\zeta^{-\alpha+1}+(\alpha-1)\zeta^{-\alpha} \right]}{a\langle\tau\rangle t^{\alpha-1}t_a^{\alpha}(1-\alpha)},
\end{aligned}
\end{equation}
which leads to Eq.~\eqref{Pxtta_ulls} in the main text.
\section{MOMENTS OF THE POSITION obtained from rare fluctuations}\label{appD}
\begin{figure}[htp]
  \centerline{\includegraphics[width=22pc]{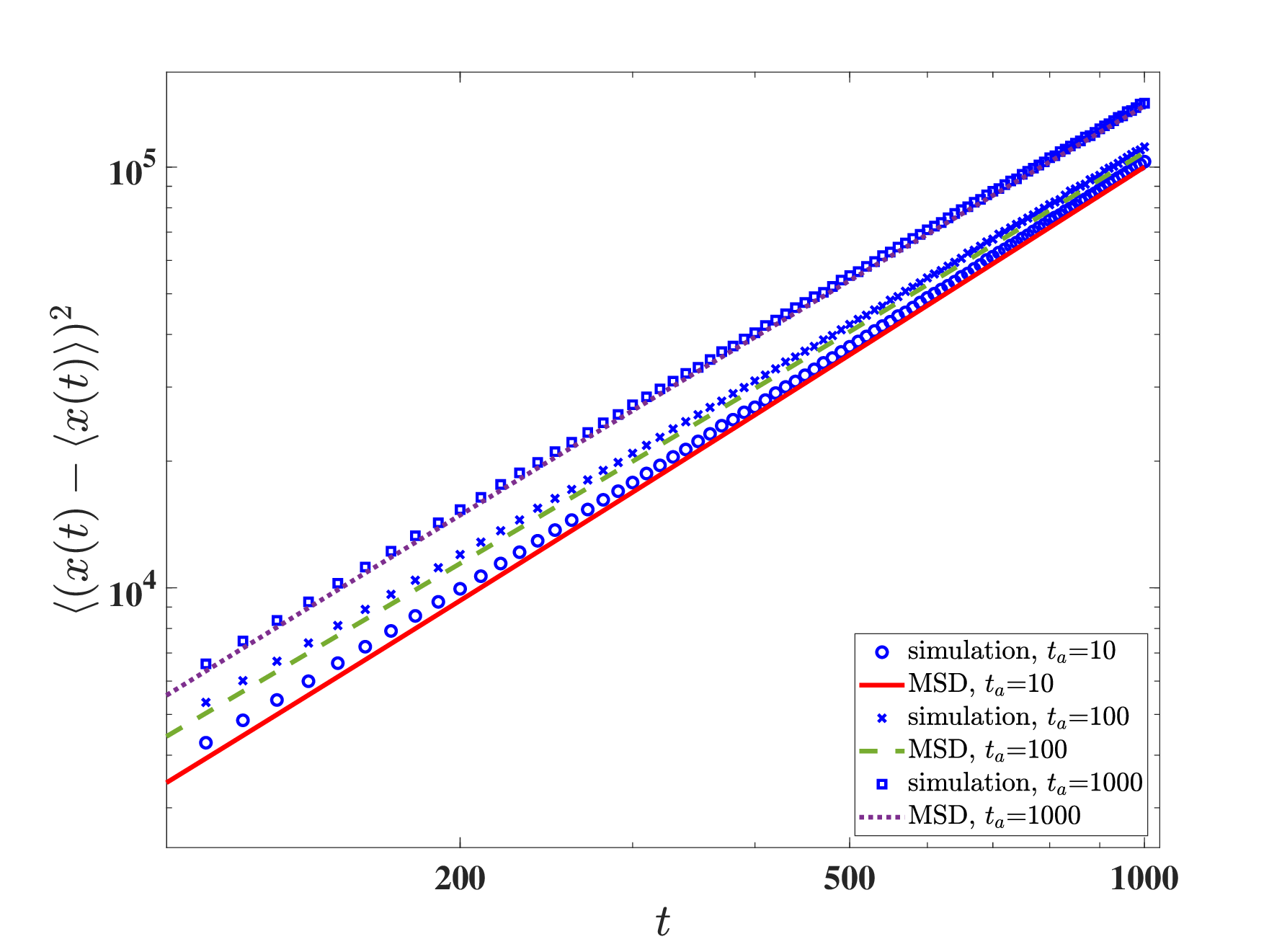}}
  \caption{The MSD for various values of $t_a$. The symbols are simulation results obtained from $10^7$  realization. The theory value is calculated from Eq.~\eqref{MSDWEAK} with different $t_a$. Here $\alpha=3/2, a=1, \sigma=1, t=1000, \tau_0=0.1$ and $t_a=10, 100, 1000$.}
  \label{fig:msd}
\end{figure}

Now, we show the derivation process of the MSD for ACTRW using rare fluctuations. Based on Eq.~\eqref{endPxtta1}, we have
\begin{widetext}
\begin{equation}\label{MSDrARE}
    \begin{aligned}
        \Big<\left(x-\dfrac{at}{\langle\tau\rangle}\right)^2\Big>=&\int_{-\infty}^{\infty}\left(x-\dfrac{at}{\langle\tau\rangle}\right)^2P(x,t_a,t)\mathrm{d}x\\
        \sim&\dfrac{(at)^2}{\langle\tau\rangle^2}\dfrac{\tau_0^{\alpha}\left[(t+t_a)^{1-\alpha}-t^{1-\alpha}\right]}{\langle\tau\rangle(1-\alpha)}+\dfrac{(at)^2}{\langle\tau\rangle^2}\int_{0}^{1}\frac{\tau_0^{\alpha}}{\langle\tau\rangle t^{\alpha-1}}\left[\dfrac{(2-\alpha)}{\zeta^{\alpha-2}}+\dfrac{\alpha}{\zeta^{\alpha-1}}-\zeta^2(\zeta+\dfrac{t_a}{t})^{-\alpha}\right]\mathrm{d}\zeta\\
        \sim& \dfrac{2a^2 b_{\alpha}}{\langle\tau\rangle^3\Gamma(4-\alpha)}\left[t^{3-\alpha}+(t+t_a)^{3-\alpha}-t_a^{3-\alpha}-(3-\alpha)(t+t_a)^{2-\alpha}t\right].
    \end{aligned}
\end{equation}
\end{widetext}
Eq.~\eqref{MSDrARE} is plotted in  Fig.~\eqref{fig:msd} and consistent with simulations.
For the ACTRW model, the non-moving particles contribute to the MSD, which differs from the CTRW model as described by large deviation theory \cite{Wang2019Transport}.
%$\langle\zeta^2\rangle =\int_{0}^{1}\frac{\tau_0^{\alpha}}{\langle\tau\rangle t^{\alpha-1}}\zeta^2\left[(2-\alpha)\zeta^{-\alpha}+\alpha\zeta^{-1-\alpha}-(\zeta+\frac{t_a}{t})^{-\alpha}\right]\mathrm{d}\zeta.$
Using the relationship between moments and the characteristic function, the asymptotic behavior of $\langle x(t)\rangle$ and $\langle x^2(t)\rangle$ follows
\begin{equation}\label{x1}
    \langle x(t)\rangle\sim \dfrac{a}{\langle\tau\rangle}t+\dfrac{a b_{\alpha}}{\langle\tau\rangle^2\Gamma(3-\alpha)}\left[(t+t_a)^{2-\alpha}-t_a^{2-\alpha}\right]
\end{equation}
and
\begin{equation}\label{x2}
\begin{aligned}
    \langle x^2(t)\rangle&\sim \dfrac{a^2}{\langle\tau\rangle^2}t^2+\dfrac{2a^2 b_{\alpha}}{\langle\tau\rangle^3\Gamma(4-\alpha)}[t^{3-\alpha}\\
  &~~~~~+(t+t_a)^{3-\alpha}\left.-t_a^{3-\alpha}-(3-\alpha)t_a^{2-\alpha}t\right],
\end{aligned}
\end{equation}
respectively. Thus, the MSD obeys
\begin{equation}\label{MSDdREI}
    \begin{aligned}
        \langle(x(t)-&\langle x(t)\rangle)^2=\langle x^2(t)\rangle-\langle x(t)\rangle^2\\
        &\sim \dfrac{2a^2 b_{\alpha}}{\langle\tau\rangle^3\Gamma(4-\alpha)}\left[t^{3-\alpha}+(t+t_a)^{3-\alpha}\right.\\
        &~~~~~~~\left.-t_a^{3-\alpha}-(3-\alpha)(t+t_a)^{2-\alpha}t\right].
    \end{aligned}
\end{equation}
It can be seen that Eq.~\eqref{MSDdREI} is the same as Eq.~\eqref{MSDrARE}.

\end{appendices}

\bibliographystyle{prestyle}
\bibliography{wenxian}

\end{document}